\documentclass[a4paper]{article}
\pdfoutput=1

\usepackage{jcappub} 

\usepackage[T1]{fontenc} 
\usepackage[utf8]{inputenc}
\usepackage{amsmath,amsfonts,amssymb}

 \usepackage{soul}
\usepackage{graphicx}
\usepackage{wrapfig}
\usepackage{subfig}
\usepackage{feynmp-auto}
\usepackage{comment}
\usepackage[normalem]{ulem}

\usepackage{hyperref}
\usepackage{tikz-feynman} 
\usepackage{cancel} 
\usepackage[dvipsnames]{xcolor}
\usepackage{bm}

\definecolor{ForestGreen}{rgb}{0.13, 0.55, 0.13}

\newcommand\Tstrut{\rule{0pt}{2.6ex}}         

\title{Sterile neutrino dark matter in conformal Majoron models}

\author[a]{Jo\~ao~Gon\c{c}alves,}
\author[b]{Danny Marfatia,}
\author[c,a]{Ant\'onio~P.~Morais}
\author[a,c,f]{Vinícius Oliveira}
\author[f]{and Roman Pasechnik}

\affiliation[a]{Laborat\'{o}rio de Instrumenta\c{c}\~{a}o e F\'{i}sica Experimental de Part\'{i}culas (LIP), Universidade do Minho, 4710-057 Braga, Portugal}
\affiliation[b]{Department of Physics and Astronomy, University of Hawaii at Manoa, Honolulu, HI 96822, USA}
\affiliation[c]{Departamento de F\'{i}sica, Escola de Ci\^{e}ncias, Universidade do Minho, 4710-057 Braga, Portugal}
\affiliation[d]{Departamento de F\'{i}sica da Universidade de Aveiro, Campus de Santiago, 3810-183 Aveiro, Portugal.}
\affiliation[f]{Department of Physics, Lund University, 221 00 Lund, Sweden}

\emailAdd{jpedropino@ua.pt}
\emailAdd{dmarf8@hawaii.edu}
\emailAdd{amorais@fisica.uminho.pt}
\emailAdd{viniciuslbo@lip.pt}
\emailAdd{roman.pasechnik@fysik.lu.se}

\abstract{
We study sterile neutrino dark matter (DM) in a classically conformal $\mathrm{U(1)}^\prime$ extension of the Standard Model with three right-handed neutrinos and a Majoron-like singlet scalar that generate the observed pattern of active neutrino masses and mixing via the type-I seesaw mechanism.  Working in the regime of strongly suppressed active-sterile mixing, we show that the observed DM abundance can be produced through freeze-in from feeble interactions mediated by the heavy $Z^\prime$ and the conformal scalar. We solve the Boltzmann equation for the nonthermal phase-space distribution and confront the scenario with Lyman-$\alpha$ and strong gravitational lensing data by computing the matter power spectrum. For keV-mass sterile neutrinos we identify the viable parameter space consistent with the inferred relic density and structure-formation and X-ray bounds, including regions compatible with a tentative 3.5~keV line. If a second sterile state is long-lived, late decays can realize a two-component setup that alleviates the $S_8$ tension. However, explanations of the 3.5~keV line and the $S_8$ tension are incompatible with the strong-lensing limit from JWST if the neutrinos saturate the relic abundance. In a highly fine-tuned variant of the model, the 220~PeV KM3NeT event can also be explained by invoking the decay of a superheavy sterile neutrino.}

\begin{document}

\maketitle
\flushbottom

\section{Introduction}

Sterile neutrinos are well-motivated dark matter (DM) candidates that span a wide range of masses and have a variety of
production histories; see Ref.~\cite{Boyarsky:2018tvu} for a review. In the keV mass window,
a sterile neutrino $N$ is intrinsically unstable due to its mixing with active neutrinos, and can 
produce observable X-ray signals through the radiative decay $N\to \nu\gamma$~\cite{Pal:1981rm}. 
This motivates dedicated line searches with current and forthcoming X-ray missions such as
\emph{Athena}~\cite{Barret:2018qft}, \emph{eROSITA}~\cite{eROSITA:2012lfj} and 
\emph{XRISM}~\cite{XRISMScienceTeam:2020rvx}.
In fact, a weak feature at $E_\gamma\simeq 3.5~\mathrm{keV}$ reported in stacked
cluster data and in observations of individual objects -- including the Perseus cluster and M31 -- triggered sustained
interest in a sterile neutrino interpretation with $m_N\simeq 7~\mathrm{keV}$ and small active-sterile
neutrino mixing $\sin^2 2\theta_\mathrm{eff} \sim 10^{-10}$~\cite{Bulbul:2014sua,Boyarsky:2014jta}. A signal near the
same energy has also been observed in the galactic center of the Milky Way in deep \emph{XMM-Newton}
exposures~\cite{Stiele_2011}.

It is well known that light DM candidates produced via thermal freeze-out in the early Universe 
typically suffer from an overabundance problem. 
Perhaps the simplest nonthermal production mechanism for sterile neutrino DM is nonresonant production
via active-sterile oscillations in the thermal plasma, the so-called Dodelson-Widrow (DW)
mechanism~\cite{Dodelson:1993je,Dolgov:2000ew}. However, with this mechanism, the inferred DM relic abundance cannot be saturated by sterile neutrinos with masses below 41~keV (at 95\%~C.L.),  a conclusion drawn from a combination of X-ray~\cite{Yuksel:2007xh,Boyarsky:2007ge,
Loewenstein:2009cm,Horiuchi:2013noa,Neronov:2016wdd,Perez:2016tcq,Roach:2019ctw,Foster:2021ngm}
and Lyman-$\alpha$ forest data~\cite{Irsic:2017ixq,Abazajian:2017tcc,Baur:2017stq}. The latter are sensitive to the phase-space distribution of sterile neutrinos and constrain their relic abundance for masses between 1~keV to 8~keV~\cite{Boyarsky:2008ju,Boyarsky:2008xj,Baur:2017stq}. Recently, strong gravitational lensing has been used to probe the DM free-streaming length and set 95\% C.L. lower limits on the DM mass between 6.1~keV and 8.4~keV if the DM saturates the relic abudance~\cite{Gilman:2025fhy,Keeley:2025oig,Keeley:2024brx,Gilman:2026uvq}.

The status of the 3.5~keV line mentioned above has been extensively scrutinized, including with follow-up searches 
and alternative (astrophysical/instrumental) explanations~\cite{Boyarsky:2018tvu,Ruchayskiy:2015onc,Urban:2014yda,
Franse:2016dln,Neronov:2016wdd,Perez:2016tcq,Cappelluti:2017ywp,XRISM:2025lzv,Dessert:2023fen}. Recent high-resolution spectroscopy continues to sharpen constraints on putative unidentified lines. For example, a stacked \emph{XRISM}/Resolve analysis of ten galaxy clusters reports no statistically significant novel features in the $2.5~\mathrm{keV}-15~\mathrm{keV}$ window and sets upper limits on the decay interpretation of the $3.5~\mathrm{keV}$ line~\cite{XRISM:2025lzv}. Furthermore, Ref.~\cite{Dessert:2023fen} finds no significant evidence for the line in attempts to reproduce previous analyses of various datasets, and instead finds evidence for background mismodeling in several analyses. Rather than address the controversy, in a part of our study we take the optimistic view that claims for the $3.5~\mathrm{keV}$ line are legitimate. Since the DW mechanism is now strongly constrained, we explore a DM freeze-in production scenario. We focus on an extremely suppressed active-sterile mixing, $\sin^2 2\theta_{\rm eff} < 10^{-10}$ in which DW production is negligible. 

We study sterile neutrino DM in a class of \emph{classically conformal} gauged $\mathrm{U(1)}^\prime$ 
extensions of the Standard Model (SM). Three sterile neutrinos are required for anomaly cancellation, and neutrino masses are generated via a type-I seesaw mechanism~\cite{Oda:2015gna}.
The conformal setup triggers spontaneous symmetry breaking radiatively via the Coleman-Weinberg 
mechanism, and the $\mathrm{U(1)}^\prime$ breaking scale generates Majorana masses for three SM-singlet 
neutrinos. The lightest sterile state, denoted $N_1$, can have feeble portal interactions and play the role of a nonthermal DM candidate.
In the keV domain, we compute the freeze-in abundance and 
apply small-scale structure constraints by propagating the resulting nonthermal 
phase-space distribution through a Boltzmann solver for cosmological perturbations. We assess the impact of this distribution on the matter power spectrum, confronting it with Lyman-$\alpha$ and strong-lensing constraints. We also 
determine the parameter regions where radiative decays of $N_1$ yield an observable X-ray 
line. 

If the additional sterile neutrino states are in the keV mass range and are long lived, they can have important cosmological implications. We consider a scenario with two keV-mass sterile neutrinos, in which the heavier neutrino, $N_2$, can decay to $N_1$. In this setup, $N_1$ receives a small momentum kick sufficient for it to freestream and suppress power on small scales. This then provides a possible connection to a persistent mild discrepancy between the amplitude of late-time matter fluctuations inferred 
from early-Universe observations (most notably Planck) and that preferred by low-redshift large-scale structure probes~\cite{Porredon:2025xxv,Semenaite:2025ohg,Lange:2025rsx}; the latter measure a smaller value than does Planck. This is widely known as the $S_8$ tension, where
$S_8 \equiv \sigma_8 \sqrt{\Omega_m/0.3}$,
with $\sigma_8$ denoting the variance of matter fluctuations on $8~\mathrm{Mpc}/h$ scales 
and $\Omega_m$ the present matter density parameter; see e.g.~\cite{Abdalla:2022yfr} for a review. {Extant resolutions of this tension with decaying DM rely on effective theory approaches or toy models~\cite{Fuss:2024dam,Astros:2025aan}. Our framework is ultraviolet complete and successfully accounts for the observed pattern of neutrino masses and mixings.

Thus, our approach ties together several aspects of keV-mass sterile neutrino DM in generic conformal Majoron models that explain active neutrino oscillation data: freeze-in production of DM and its relic abundance, constraints from Lyman-$\alpha$, strong-lensing and X-ray data, and a multi-component decaying DM explanation of the $S_8$ tension. Additionally, we briefly comment on how, in a different region of the parameter space, these models accommodate a very heavy 
sterile neutrino that may explain the recently reported $\sim 200$~PeV KM3NeT event~\cite{KM3NeT:2025npi} via its decay~\cite{Kohri:2025bsn}.

The article is organized as follows. In Section~\ref{sec:theory} we introduce the classically conformal
$\mathrm{U(1)^\prime}$ Majoron-type model and specify its neutrino, scalar, and 
gauge sectors. In Section~\ref{sec:stability} we discuss the cosmological stability 
and X-ray constraints on keV sterile neutrinos and summarize the relevant decay channels and lifetimes. 
In Section~\ref{sec:freezein} we present the freeze-in production mechanism and the computation 
of the relic abundance. In Section~\ref{sec:LymanAlpha} we confront the resulting nonthermal 
DM spectrum with Lyman-$\alpha$ forest and strong-lensing constraints. The phenomenology of the $3.5$~keV X-ray 
line and its interpretation in terms of $N_1\to\nu\gamma$ is addressed in Section~\ref{sec:35keVline}. In Section~\ref{sec:S8} we demonstrate how a multi-component (decaying) DM realization resolves the $S_8$ tension. In Section~\ref{sec:EeV-scale} we comment on how  the decay of a super-heavy 
sterile neutrino can explain the KM3NeT event. We summarize our findings  
in Section~\ref{sec:conclusions}.

\section{Theoretical setup}
\label{sec:theory}

We work within a class of classically conformal, anomaly-free gauged $\mathrm{U(1)}^{\prime}$ extensions of 
SM supplemented by three right-handed neutrinos $N_I$ ($I=1,2,3$) required by anomaly 
cancellation, and a complex SM-singlet scalar $\sigma$ that generate neutrino masses~\cite{Chikashige:1980qk,Chikashige:1980ui,Gelmini:1980re}. Classical conformal invariance 
forbids explicit mass terms in the tree-level Lagrangian; all mass scales arise from radiative symmetry 
breaking via the Coleman-Weinberg mechanism and the induced vacuum expectation values (VEVs). The model 
setup and the one-loop treatment of the scalar sector follow Ref.~\cite{Goncalves:2024lrk}.

Throughout, we parametrize the flavor-universal $\mathrm{U(1)}^{\prime}$ charges in terms of the charges $x_{\mathcal H}$ and $x_{\sigma}$ of the SM Higgs doublet $\mathcal H$ 
and $\sigma$, respectively. The remaining charge assignments are fixed by gauge 
invariance of the Yukawa sector and by anomaly cancellation. The resulting field content and charges 
are provided in Table~\ref{tab:charges}.
\begin{table}[t]
	\centering
	\begin{tabular}{c|c|c|c|c}
		\textbf{Field} & $\mathbf{U(1)^{\prime}}$ & $\mathbf{SU(3)_\text{c}}$ & $\mathbf{SU(2)_\text{L}}$ & $\mathbf{U(1)_\text{Y}}$  \\ \hline\Tstrut
		$Q$ & $\frac{1}{3} x_{\mathcal H} + \frac{1}{6} x_{\sigma}$ & $\bm{3}$ & $\bm{2}$ & $1/6$  \\[0.5em]
		$u_R$ & $\frac{4}{3} x_{\mathcal H} + \frac{1}{6} x_{\sigma}$ & $\bm{3}$ & $\bm{1}$ & $2/3$ \\[0.5em]
  		$d_R$ & $-\frac{2}{3} x_{\mathcal H} + \frac{1}{6} x_{\sigma}$ & $\bm{3}$ & $\bm{1}$ & $-1/3$ \\[0.5em]
		$L$ & $ -x_{\mathcal H} - \frac{1}{2} x_{\sigma}$ & $\bm{1}$ & $\bm{2}$ & $-1/2$ \\[0.5em]
        $e_R$ & $ -2x_{\mathcal H} - \frac{1}{2} x_{\sigma}$ & $\bm{1}$ & $\bm{1}$ & $-1$ \\[0.5em]
		$\mathcal{H}$ & $x_{\mathcal H}$ & $\bm{1}$ & $\bm{2}$ & $1/2$ \\[0.5em]
        $N$ & $-\frac{1}{2} x_{\sigma}$ & $\bm{1}$ & $\bm{1}$ & $0$ \\[0.5em]
		$\sigma$ & $x_{\sigma}$ & $\bm{1}$ & $\bm{1}$ & $0$ \\ \hline
	\end{tabular}
	\caption{Field content and gauge charges in the conformal $\mathrm{U(1)}^{\prime}$ framework. All
    $\mathrm{U(1)}^{\prime}$ charges are expressed in terms of the Higgs charge $x_{\mathcal H}$ and singlet 
    charge $x_{\sigma}$; the other assignments follow from gauge invariance and anomaly cancellation.}
	\label{tab:charges}
\end{table}

\subsection{Neutrino sector}

The neutrino Yukawa Lagrangian reads~\cite{Chikashige:1980qk,Chikashige:1980ui,Gelmini:1980re} 
\begin{equation}
\mathcal{L}_{\nu} \supset  y_{\nu}^{ij}\,\overline{L_i}\,\tilde{\mathcal H}\,N_j + y_{\sigma}^{I}\,\overline{N_I^{c}}\,N_I\,\sigma + \text{h.c.}\,,
\label{eq:Lnu}
\end{equation}
where $L_i = (\nu_{\mathrm{L}i}, e_{\mathrm{L}i} )^\mathrm{T}$, $i,j=1,2,3$, is the SM lepton doublet, and  
$\tilde{\mathcal H} = i\sigma_2\,\mathcal H^{*}$. After symmetry breaking,
\begin{equation}
\langle \mathcal H \rangle = \frac{1}{\sqrt{2}}
\begin{pmatrix}
0 \\ v
\end{pmatrix}\,,\qquad 
\langle \sigma \rangle = \frac{v_{\sigma}}{\sqrt{2}}\,,
\end{equation}
where $v\simeq 246~\text{GeV}$ and $v_{\sigma}$ is generated radiatively. 
In the basis $(\nu_L,\,N^{c})$ the $6\times6$ Majorana mass matrix is
\begin{equation}
\bm{M_\nu} = \left( \begin{array}{cc}
0        &  \tfrac{v}{\sqrt{2}} \,\bm{y_\nu}^{\mathrm{T}} \\  
\tfrac{v}{\sqrt{2}} \,\bm{y_\nu}  & \tfrac{v_\sigma}{\sqrt{2}}\,\bm{y_\sigma}
\end{array} \right)\,.
\label{eq:Mnu}
\end{equation}
Assuming a standard type-I seesaw hierarchy $\|\tfrac{v}{\sqrt{2}}\bm{y_\nu}\|\ll \|\tfrac{v_\sigma}{\sqrt{2}}\bm{y_\sigma}\|$, the light neutrino mass matrix is
\begin{equation}
\bm{m}_\nu \simeq \frac{1}{\sqrt{2}}\,\frac{v^2}{v_{\sigma}}\,\bm{y_\nu}^{\mathrm{T}}\,\bm{y_\sigma}^{-1}\,\bm{y_\nu}\,,
\label{eq:nu-light}
\end{equation}
whose eigenvalues correspond to the three active neutrinos $m_1,m_2,m_3$. The heavy neutrino mass matrix is
\begin{equation}
\bm{M_{N}} \simeq \frac{v_{\sigma}}{\sqrt{2}}\,\bm{y_\sigma}\,,
\label{eq:heavyN}
\end{equation}
with eigenstates denoted by $N_I$ with masses $M_{N_I}$.  We identify 
the lightest sterile state $N_1$ as the DM candidate. In the keV-scale scenarios studied below we take
$M_{N_1}\in[1,100]~\mathrm{keV}$, while $M_{N_2}$ and $M_{N_3}$ are treated as free parameters 
(with special attention to the case where $N_2$ is also long-lived).

For parameter scans it is convenient to parameterize the Dirac Yukawa matrix using 
the Casas-Ibarra construction~\cite{Casas:2001sr}. The type-I seesaw relation \eqref{eq:nu-light} can be written as
\begin{equation}
\bm{m}_\nu \simeq \bm{y}_\nu^{\mathrm{T}}\,\bm{\Sigma}\,\bm{y}_\nu\,,
\qquad
\bm{\Sigma}\equiv \frac{1}{\sqrt{2}}\frac{v^2}{v_{\sigma}}\,\bm{y}_{\sigma}^{-1}
= \frac{v^2}{2}\,\bm{M}_N^{-1}\,,
\label{eq:Sigma-def}
\end{equation}
with \cite{Cordero-Carrion:2019qtu}
\begin{equation}
\bm{y}_\nu = i\,\bm{\Sigma}^{-1/2}\,\bm{R}\,\bm{D}_{\sqrt{m}}\,\bm{U}_{\rm PMNS}^{\dagger}\,,
\label{eq:Casas-Ibarra-Sigma}
\end{equation}
where $\bm{D}_{\sqrt{m}}=\mathrm{diag}(\sqrt{m_1},\sqrt{m_2},\sqrt{m_3})$, $\bm{U}_{\rm PMNS}$ 
is the Pontecorvo-Maki-Nakagawa-Sakata (PMNS) mixing matrix, and $\bm{R}$ is a complex orthogonal matrix that
satisfies $\bm{R}^{\mathrm{T}}\bm{R}=\mathbf{1}$. In the heavy neutrino mass basis, $\bm{\Sigma}$ 
is diagonal with eigenvalues, $\Sigma_I=v^2/(2M_{N_I})$. The matrix $\bm{R}$ 
can be parameterized as a product of three 3D complex rotations $\mathbf{R}_{iI}$ in each $iI$-plane, i.e.,
$\bm{R} = \mathbf{R}_{23}(z_1)\,\mathbf{R}_{13}(z_2)\,\mathbf{R}_{12}(z_3)$, with complex angles $z_k$. 
This parameterization guarantees compatibility with the measured active neutrino masses and mixing angles while 
allowing a broad exploration of the sterile sector. As a phenomenological input, we utilize the results of NuFIT~\cite{Esteban:2024eli} for the neutrino mass differences and PMNS mixing angles 
and apply the cosmological bound on the sum of active neutrino masses, 
$\sum_{i=1,2,3} m_i < 0.12~\mathrm{eV}$~\cite{Planck:2018vyg}. 

\subsection{Scalar sector}

The classically conformal tree-level scalar potential reads 
 \begin{equation}\label{eq:tree_largemh2}
     V_{0}(\mathcal{H},\sigma) = \lambda_h(\mathcal{H}^{\dagger}\mathcal{H})^2 + \lambda_{\sigma}(\sigma^{\ast}\sigma)^2 + \lambda_{\sigma h}(\mathcal{H}^{\dagger}\mathcal{H})(\sigma^{\ast}\sigma) \,.
 \end{equation}
$\mathcal{H}$ and $\sigma$ develop 
nonzero VEVs upon radiative symmetry breaking.
In a conformal setting, one scalar degree of freedom is massless at tree level 
along the flat direction and can be identified as the Goldstone boson of spontaneously broken 
scale invariance. Radiative corrections explicitly break the scale invariance via 
the Coleman–Weinberg (CW) mechanism by lifting the flat direction 
of the tree-level potential, thus generating a nonzero 
pseudo-Goldstone mass term.

Expanding the fields about the vacuum as
\begin{equation}
\mathcal H = \frac{1}{\sqrt{2}}\begin{pmatrix}0\\ v+h\end{pmatrix}\,, \qquad 
\sigma=\frac{1}{\sqrt{2}}\,(v_{\sigma}+s)\,e^{iJ/v_{\sigma}}\,,
\end{equation}
the spectrum contains a physical CP-even state dominantly aligned with the SM Higgs $h_1$, 
a singlet-like CP-even state $h_2$, and a CP-odd Majoron $J$ providing the longitudinal 
mode of the $Z^{\prime}$ boson. The CP-even fields $(h,s)$ are related to mass eigenstates $(h_1,h_2)$ 
by an orthogonal rotation,
\begin{equation}
\begin{pmatrix} h_1 \\ h_2 \end{pmatrix}=\mathcal O\,\begin{pmatrix} h \\ s \end{pmatrix}\,, \qquad
\mathcal{O} =
\begin{pmatrix}
\cos \alpha & \sin \alpha \\
-\sin \alpha & \cos \alpha
\end{pmatrix}\,.
\label{eq:scalar-mix}
\end{equation}
In our framework, we identify $h_1$ with the observed SM-like Higgs boson, with mass $M_{h_1}\simeq 125.11~\mathrm{GeV}$, which includes significant one-loop corrections. The heavier scalar $h_2$ is aligned with the pseudo-Goldstone boson of scale symmetry 
breaking with mass generated radiatively at one loop level.

We follow Ref.~\cite{Goncalves:2024lrk} for the one-loop effective potential, tadpole conditions, 
and mass renormalization. First, we evaluate the tadpole equations and the scalar mass spectrum at 
the one-loop level by including the CW potential $V_\mathrm{CW}$, in the total potential $V_\mathrm{tot} = V_0 + V_\mathrm{CW}$, and 
then determine the scalar mass spectrum by incorporating finite-momentum corrections due to self-energies.
This procedure allows us to determine $\lambda_\sigma$ and $\lambda_{\sigma h}$ through the one-loop 
tadpole conditions, and $\lambda_h$ and $v_\sigma$ from the one-loop corrected scalar masses, 
leaving the mass of the heavy scalar $M_{h_2}$ as the only 
free parameter in the scalar sector. Higgs signal-strength measurements and direct searches 
for additional scalars constrain $|\sin \alpha| < 0.3$~\cite{ATLAS:2022vkf}. The phenomenologically viable parameter space 
relevant for freeze-in DM corresponds to large $\mathrm{U(1)}^\prime$ breaking scales, 
$v_\sigma > 1~\mathrm{TeV}$, with a strongly suppressed scalar mixing angle, 
$|\sin\alpha|\lesssim 10^{-2}$. The coupling of $h_1$ to light $N_1$  induced by mixing $\propto M_{N_1}\sin\alpha/v_{\sigma}$ and yields invisible Higgs decay with width
\begin{equation}
\Gamma_{h_1\to \bar N_1 N_1}
=\frac{M_{N_1}^2}{8\pi\,v_\sigma^2}\,\sin^2\alpha\;M_{h_1}
\left(1-\frac{4M_{N_1}^2}{M_{h_1}^2}\right)^{3/2}\,.
\end{equation}
We will find this to be negligible compared 
to the SM Higgs boson width.

\subsection{Gauge sector}

The $\mathrm{U(1)}^{\prime}$ gauge field $B^\prime_\mu$ couples to the $\mathrm{U(1)}^{\prime}$ current 
with coupling $g_L$, and gauge invariance allows kinetic mixing between the
$\mathrm{U(1)}^{\prime}$ field strength $B^\prime_{\mu\nu}$ and the hypercharge field strength $B_{\mu\nu}$. 
We parameterize the mixing at the level of covariant derivatives by
introducing an effective mixing parameter $g_{12}$~\cite{Goncalves:2024lrk}. After electroweak and $\mathrm{U(1)}^{\prime}$ breaking, the neutral gauge boson mass-squared matrix 
in the $(B_\mu,\,A^3_\mu,\,B^\prime_\mu)$ basis reads
\begin{equation}\label{eq:Mass_Matrix_V}
\mathcal{M}^{2}_V=
    \left(
\begin{array}{ccc}
 \frac{1}{4} g_2^2 v^2 \tan ^2 \theta_\mathrm{W} & -\frac{1}{4} g_2^2 v^2 \tan  \theta_\mathrm{W} & 
 \frac{1}{4} g_2 v^2 \tan  \theta_\mathrm{W} (g_{12}+2 g_L x_{\mathcal H}) \\
 -\frac{1}{4} g_2^2 v^2 \tan \theta_\mathrm{W} & \frac{1}{4}g_2^2 v^2 & -\frac{1}{4} g_2 v^2 
 (g_{12}+2 g_L x_{\mathcal H}) \\
 \frac{1}{4} g_2 v^2 \tan  \theta_\mathrm{W} (g_{12}+2 g_L x_{\mathcal H}) & 
 -\frac{1}{4} g_2 v^2 (g_{12}+2 g_L x_{\mathcal H}) & g_L^2 v_{\sigma}^2 x_{\sigma}^2+\frac{1}{4} v^2
 (g_{12}+2 g_L x_{\mathcal H})^2
\end{array} \right)\,,
\end{equation}
where $\theta_\mathrm{W}$ is the Weinberg angle and $g_2$ is the $\mathrm{SU(2)_L}$ 
gauge coupling. Thus, $g_L$, $g_{12}$, $x_\mathcal{H}$ and $x_\sigma$ are free parameters that
fully determine the gauge sector. Diagonalization yields a massless photon and two massive 
states identified with the $Z^0$ boson of the SM and a massive $Z^{\prime}$ vector 
boson; expressions for their masses can be found in Ref.~\cite{Goncalves:2024lrk}.

It is well known that light DM particles produced via  freeze-out  suffer from an overabundance problem. To avoid this issue, we focus on the regime in which the DM  never reaches thermal equilibrium, and is produced nonthermally via the freeze-in mechanism. We are interested in the limit $g_L,g_{12} \ll 1$ and $v_{\sigma}\gg v$, in which case
\begin{equation}
M_{Z}^2\simeq \frac{1}{4}g_2^2 (1+\tan^2\theta_\mathrm{W})v^2\,,\qquad
M_{Z^{\prime}}^2\simeq (g_L x_{\sigma}v_{\sigma})^2\,.
\label{eq:Zmass-approx}
\end{equation}
We require $M_{Z^{\prime}}\gtrsim 5~\mathrm{TeV}$ to satisfy bounds from the LHC because the U(1)$^\prime$ charges in Table~\ref{tab:charges} are generation-independent~\cite{ATLAS:2020lks,ATLAS:2020tre,CMS:2024vhy}. We therefore take a conservative approach and impose $5~\mathrm{TeV}$ as the lower bound in our analysis.
In conformal $\mathrm{U(1)}^{\prime}$ models, $\lambda_{\sigma}\sim (g_Lx_\sigma)^4$~\cite{Khoze:2014xha}, 
so that small $g_L$ simultaneously helps satisfy collider bounds and ensures the hierarchy
$M_{h_2} < M_{Z^{\prime}}$ which is needed in the DM production analysis below.

\section{Stability and X-ray constraints on keV sterile neutrinos}
\label{sec:stability}

The DM candidate $N_1$ is not absolutely stable. It inherits weak interactions
through its mixing with the active neutrinos. Let $U$ denote the unitary matrix that 
diagonalizes the neutrino mass matrix in Eq.~\eqref{eq:Mnu}, such that the active flavor states 
contain a small admixture of sterile states. For the lightest sterile neutrino $N_1$, 
we define an effective active-sterile mixing parameter,
\begin{equation}
\sin^2 \theta_{\rm eff} \equiv \sum_{\alpha=e,\mu,\tau} |U_{\alpha (I=1)}|^2 \ll 1 \,,
\label{eq:thetaeff-def}
\end{equation}
which determines the primary SM-induced decay channels. In the seesaw regime, 
$U_{\alpha (I=1)}\simeq (m_D)_{\alpha (I=1)}/M_{N_1}$. 
For keV $N_1$, $\theta_{\rm eff}$ is tiny because the Dirac neutrino 
mass scale $m_D=vy_\nu/\sqrt2$. This enables
the $N_1$ lifetime to exceed the age of the Universe, 
$\tau_0 \simeq 13 \times 10^9$ yr. 

Light $N_1$ mainly decay via the two-body $N_1\to\nu\gamma$ and 
the three-body invisible $N_1\to\nu\nu\bar\nu$ channels~\cite{Helo:2010cw}. 
The one-loop contribution to the radiative $N_1\to\nu\gamma$ decay width 
is induced by $W$-exchange and given by~\cite{Pal:1981rm,Barger:1995ty,Drewes:2016upu}
\begin{equation}
\Gamma_{N_1\to\nu\gamma}
\simeq \frac{9\,\alpha_{\rm EM}\,G_F^2}{256\,\pi^4}\,
\sin^2\theta_{\rm eff}\,M_{N_1}^5\simeq 5.5\times10^{-22}\,
\sin^2\theta_{\rm eff}
\left(\frac{M_{N_1}}{1~\mathrm{keV}}\right)^5~\mathrm{s}^{-1}\,,
\label{eq:rad_gammadecay}
\end{equation}
where $\alpha_\mathrm{EM}$ is the fine-structure constant and $G_F$ is the Fermi constant. 
In our model, an additional contribution to 
$\Gamma_{N_1\to\nu\gamma}$ arises due to kinetic mixing between $\mathrm{U(1)_Y}$ and $\mathrm{U(1)}^\prime$. 
However, this contribution is subleading in the 
regime of small $g_{12},g_L\ll 1$ despite being generated at tree level. Finally, 
the mass mixing-induced three-body decay proceeds via an off-shell $Z^0$ boson and yields \cite{Barger:1995ty}
\begin{equation}
\Gamma_{N_1\to3\nu} \simeq \frac{G_F^2}{96\,\pi^3}\,
\sin^2\theta_{\rm eff}\,M_{N_1}^5\,,
\label{eq:Gamma-3nu}
\end{equation}
which we evaluated with
\texttt{MadGraph}~\cite{Alwall:2014hca}. Requiring that the corresponding lifetime is longer than the age of the Universe gives~\cite{Dolgov:2000ew}
\begin{equation}
    \sin^2 \theta_{\rm eff} < 1.1 \times 10^{-7} \left( \frac{50 \, \text{keV}}{m_{N_1}}\right)^5\,.
\end{equation}
This decay channel is invisible and dominant compared to the radiative loop-induced 
channel:
\begin{equation}
\frac{\Gamma_{N_1\to\nu\gamma}}{\Gamma_{N_1\to3\nu}}\simeq 
\frac{27\,\alpha_{\rm EM}}{8\pi}\,\sim\,\mathcal O(10^{-2})\,.
\label{eq:ratio}
\end{equation}
Thus, the total decay width of $N_1$ can be approximated as
\begin{equation}
\Gamma_{N_1}\simeq \Gamma_{N_1\to3\nu} + \Gamma_{N_1\to\nu\gamma}\,.
\label{eq:Gamma-total}
\end{equation}

In practice, the DM stability bound translates into a bound on the decay width, 
$\Gamma_{N_1} < \tau_0^{-1} = 1.5 \times 10^{-42}~\text{GeV}$, which we enforce 
in our numerical search for viable parameter space regions.

Besides the need to guarantee a sufficiently long lifetime for $N_1$, the one-loop decay $N_1\to\nu\gamma$ leads to the production of a mono-energetic photon at $E_\gamma\simeq M_{N_1}/2$ and is therefore 
tightly constrained by X-ray line searches. For $M_{N_1}\sim\mathcal O(\mathrm{keV})$, the corresponding 
bounds typically require $\sin^2 2\theta_{\rm eff}\lesssim 10^{-10}$--$10^{-11}$, depending on 
the dataset. Note that these constraints already imply that lifetimes $\tau_{N_1}\equiv\Gamma_{N_1}^{-1}$ are well above cosmological timescales. A compilation of X-ray constraints sets $\sim 2\sigma$ upper bounds in the range $\sin^2 2\theta_{\rm eff} \in [10^{-15},10^{-10}]$ for $M_{N} \in [1,100]~\mathrm{keV}$~\cite{Boyarsky:2007ge,Horiuchi:2013noa,Roach:2019ctw,Foster:2021ngm}, which in turn restricts the $z_i$ angles in the $\bm{R}$ 
matrix of Eq.~\eqref{eq:Casas-Ibarra-Sigma} to have magnitudes $|z_i| \lesssim 10^{-4}$. 

\section{Freeze-in production and relic abundance of keV dark matter}
\label{sec:freezein}

Since $\sin^2 2\theta_{\rm eff} \lesssim
10^{-10}$, the abundance of DM produced through DW can be neglected.
 In addition, self-interactions are highly suppressed 
due to the large masses of the mediators $h_2$ and $Z^\prime$. In this regime, 
the interactions connecting the $\mathrm{U(1)}^{\prime}$ sector to the SM are so feeble that 
the lightest sterile neutrino $N_1$ never attains thermal equilibrium with the SM plasma, 
and its cosmological abundance is produced gradually through the so-called 
\emph{freeze-in} mechanism~\cite{Hall:2009bx}. In this scenario, initially no dark particles are present, and rare scatterings among bath particles continuously generate a small population of $N_1$ states until the temperature of the Universe falls well below the relevant reaction  thresholds. In what follows, we investigate the conditions needed to reproduce  the observed relic abundance assuming, for simplicity, no initial abundance of DM while ensuring  stability and consistency with  current cosmological and astrophysical bounds.

Assuming a standard radiation-dominated cosmology and an SM thermal bath in equilibrium, the number density of a sterile species $N_I$ produced dominantly via $2\to2$ annihilation  processes is governed by the Boltzmann equation, 
\begin{equation}
\dot n_{N_I}+3Hn_{N_I}= \Gamma(\text{SM\;SM} \to N_I N_I) - \Gamma(N_I N_I \to \text{SM\;SM})\,,
\label{eq:BE-n}
\end{equation} 
where $\Gamma(\text{SM\;SM} \to N_I N_I)$ and $\Gamma(N_I N_I \to \text{SM\;SM})$ are the $N_I$ production and annihilation rates per unit volume, respectively. The Hubble rate at temperature $T$ in the radiation era is
\begin{equation}
H(T)=\sqrt{\frac{\pi^2}{90}\,g_{\rho}(T)}\,\frac{T^2}{M_{\rm Pl}} \,,
\end{equation}
 with $M_{\rm Pl}=2.44\times 10^{18}\,\mathrm{GeV}$ the reduced Planck mass, and $g_{\rho}(T)$ the effective number of energetic degrees of freedom. 

The  reaction rate per unit volume is defined  as 
\begin{align}
    \Gamma(\text{SM\;SM} \to N_I N_I) & \equiv  \langle\sigma(\text{SM\;SM} \to N_I N_I) v_r\rangle (n_\text{SM}^\text{eq})^2\,, 
\end{align}
where $\langle\sigma(\text{SM\;SM} \to N_I N_I) v_r\rangle$ denotes the thermal average of the production cross section times the relative velocity. Hence, the interaction rate is given by

\begin{align}
    \mathcal{R}(\text{SM\;SM} \to N_I N_I)
    & =
    \frac{\Gamma(\text{SM\;SM} \to N_I N_I)}{n_{\text{SM}}^{\mathrm{eq}}} \,.
\end{align}
The Maxwell-Boltzmann equilibrium number densities for the SM particles and $N_I$ are given by
\begin{equation}
n^{\rm eq }_{\rm SM}= \frac{g_{\rm SM}}{2\pi^2}M_{\rm SM}^2T\,K_2(M_{\rm SM}/T) \,, \qquad
    n^{\rm eq }_{N_I}=\frac{g_N}{2\pi^2}M_{N_I}^2T\,K_2(M_{N_I}/T) \,, \qquad 
\end{equation}
where $g_{\rm SM}$ is the number of internal degrees of freedom of the SM particle, $g_N = 2$ for Majorana fermions, and $K_r(x)$ is a modified Bessel function of the second kind of order $r$.

In the freeze-in regime, $\Gamma(N_I N_I \to \text{SM\;SM}) \ll \Gamma(\text{SM\;SM} \to N_I N_I)$, and Eq.~\eqref{eq:BE-n} reduces to
\begin{equation}
\dot n_{N_I}+3Hn_{N_I}\simeq\Gamma(\text{SM\;SM} \to N_I N_I) \,.
\label{eq:BE-n-freezein}
\end{equation}
Switching to the comoving yield $Y_{N_I}\equiv n_{N_I}/s$ and the dimensionless 
inverse temperature $x\equiv M_{N_I}/T$ leads to
\begin{align}
\frac{dY_{N_I}}{dx}& = \sqrt{\frac{8\pi^2 M_\text{Pl}^2}{45}} \frac{g_*^{1/2} M_{N_I}}{x^2} \langle \sigma(\text{SM\;SM} \to N_I N_I) v_r \rangle (Y_\text{SM}^\text{eq})^2
 \simeq\frac{s}{Hx}\,\langle\sigma(\text{SM\;SM} \to N_I N_I) v_r\rangle\,(Y_\text{SM}^{\rm eq})^2\,,
\label{eq:Boltzmann_Y}
\end{align}
where $s(T)=(2\pi^2/45)\,g_s(T)\,T^3$ is the entropy density, $g_s(T)$ is the effective number of entropic degrees of freedom, and
\begin{equation}
Y_{N_I}^{\rm eq}(x)=\frac{45}{4\pi^4}\,\frac{g_{N}}{g_s}\,x^2 K_2(x)\,, \qquad g_*^{1/2} \equiv \frac{g_s}{g_\rho^{1/2}}\left( 1 + T \frac{d g_s/dT}{3 g_s} \right)\,.
\end{equation}
The approximate equality in Eq.~\eqref{eq:Boltzmann_Y} is obtained by neglecting a subleading derivative term $\propto dg_s/dT$. Furthermore, the thermal averaged production cross section is ~\cite{Gondolo:1990dk} 
\begin{align}
\label{eq:sigmav}
     \langle \sigma(\text{SM\;SM} \to N_I N_I) v_r \rangle & =
     \frac{T}{2(2\pi)^4 \, \big(n_\text{SM}^{\rm eq}\big)^{2}}
     \int_{s_{\min}}^\infty \! ds \;
     \sigma(s)\, (s - 4 M_{\rm SM}^2)\, \sqrt{s}\, K_1(\sqrt{s}/T)\,.
\end{align}
Here, $\sigma(s)$ denotes the total $\text{SM\;SM} \to N_IN_I$ production cross section evaluated at
invariant mass squared $s$ and summed over all kinematically accessible SM final states. 
We take  $s_{\min}=\max(4M_{N_I}^2,4M_{\rm SM}^2)$ to enforce kinematic thresholds explicitly. 
The present-day relic density is obtained in terms of the asymptotic yield $Y_{N_1}^{\infty}$:
\begin{equation}
\Omega_{N_1}h^2=\frac{M_{N_1}\,s_0}{\rho_c/h^2}\,Y_{N_1}^{\infty}
\simeq 2.745\times10^{8}\left(\frac{M_{N_1}}{\mathrm{GeV}}\right)Y_{N_1}^{\infty} \,,
\label{eq:OmegaY}
\end{equation}
where $s_0 = 2891.2$ cm$^{-3}$ is the entropy density today, and $\rho_c = 1.053 \times 10^{-5} \, h^2 \, \text{GeV cm}^{-3}$ is the critical density \cite{ParticleDataGroup:2024cfk}. 

In our model, $N_1$ is produced through annihilation of SM states into $N_1N_1$ via heavy  $s$-channel mediators:  the $Z^{\prime}$ and the Majoron-like $h_2$. The dominant contributions arise from
\begin{align}
\mathrm{SM\;SM}& \longleftrightarrow N_1N_1\,,\qquad 
\mathrm{SM\;SM}=\big\{f\bar f,\;h_1h_1,\;VV,\ldots\big\}, 
\end{align}
where $f$ and $V$ denote SM fermions and gauge bosons, respectively. Their relative importance depends on $M_{Z^{\prime}}$, $M_{h_2}$, $g_L$, and the scalar-sector parameters that determine the effective $h_1h_1h_2$ interaction. In the conformal regime, $|\sin\alpha|$ is  typically tiny, so channels that need sizable $h_1$-$h_2$ mixing or $Z$-$Z^{\prime}$ mixing through  kinetic mixing are suppressed. Nevertheless, we retain all $2\to2$ contributions in the numerical analysis.  To compute the required matrix elements and cross sections, we implement the model in \texttt{SARAH}~\cite{Staub:2013tta} to generate the Feynman rules and export them to \texttt{CalcHEP}~\cite{Belyaev:2012qa}. Thermal averaging is performed using Eq.~\eqref{eq:sigmav},  and the yield in Eq.~\eqref{eq:Boltzmann_Y} is solved numerically from a reheating temperature $T_{\rm RH}=200$~TeV down to $T\ll M_{N_1}$. We have checked that, for this reheating temperature, none of the dark-sector states thermalizes with the SM bath. Throughout the scanned parameter space,  $Z^\prime$ is heavier than $T_{\rm RH}$, and its abundance is consequently Boltzmann suppressed. On the other hand, $h_2$ couples to the bath primarily through the mixing angle $\alpha$ and its interaction rate remains below the Hubble rate. The production of $N_1$ is thus dominated by the annihilation of SM particles.

Besides the $2\to2$ annihilation processes included in the freeze-in calculation, $N_1$ could in principle be produced through $Z\to\nu N_1$ and $W\to\ell N_1$.
In the electroweak plasma, however, these processes are governed by the in-medium active-sterile mixing angle. Away from an in-medium resonance, and in the regime dominated by the thermal potential of the active neutrinos $\lvert V_T\rvert\gg\Delta$, where $\Delta=M_{N_1}^2/2p$, one has $\sin^22\theta_m\simeq\sin^22\theta_{\rm eff}\,(\Delta/V_T)^2$~\cite{Abazajian:2005gj,Patwardhan:2015kga}. 
For $M_{N_1}=1$--100~keV and electroweak temperatures,
$\Delta/\lvert V_T\rvert$ is extremely small.\footnote{For the largest mass considered, $M_{N_1}=100$~keV, and
typical momenta $p\simeq 3T$ at $T\simeq M_W$, 
$\Delta \lesssim 2\times10^{-11}$~GeV.
Therefore, the potential-dominated approximation is valid if $\lvert V_T\rvert\gg 2\times10^{-11}$~GeV.
The finite-temperature potential in the plasma is much larger
than this during the epoch relevant for weak-boson decays, so that
$\Delta/\lvert V_T\rvert\ll 1$ throughout the production era.
} Consequently, $W$- and $Z$-decays contribute negligibly to
the relic abundance compared with the $2\to2$ freeze-in processes.
A possible $h_1\to\nu N_1$ channel induced by
active-sterile mixing has a very suppressed width $\propto (M_{N_1}/v)^2 \sin^2\theta_{\rm eff}$ and is
likewise irrelevant.

A necessary consistency requirement for freeze-in is nonthermalization of the sterile neutrinos 
with the SM bath. We demand that the $N_I$ production rate $\mathcal{R}(\text{SM\;SM} \to N_I N_I)$  be always below the Hubble rate, i.e., 
\begin{align}
\mathcal{R}(\text{SM\;SM} \to N_I N_I)\,<\,H(T) 
\label{eq:nontherm}
\end{align}
at all temperatures relevant to DM production. We explicitly verify Eq.~\eqref{eq:nontherm} for all viable scenarios and show $\mathcal{R}(\text{SM\;SM} \to N_I N_I)/H$ for two benchmark points with $M_{N_1}=7\,\mathrm{keV}$ and $M_{N_1}=100\,\mathrm{keV}$ in Fig.~\ref{fig:DM_Interaction_Rate}.  The purple curves represent the contribution from Higgs production, and the orange curve corresponds to production from all SM fermion pairs. 
In the left panel, the bumps in the production rates originate from the $h_1$ and $h_2$ resonances.  In the right panel, this feature and production via other SM particles are absent due to the much smaller value of $\alpha$. 

\begin{figure}[t]
   \centering
  \begin{minipage}{0.48\textwidth}
    \centering
    \includegraphics[width=\linewidth]{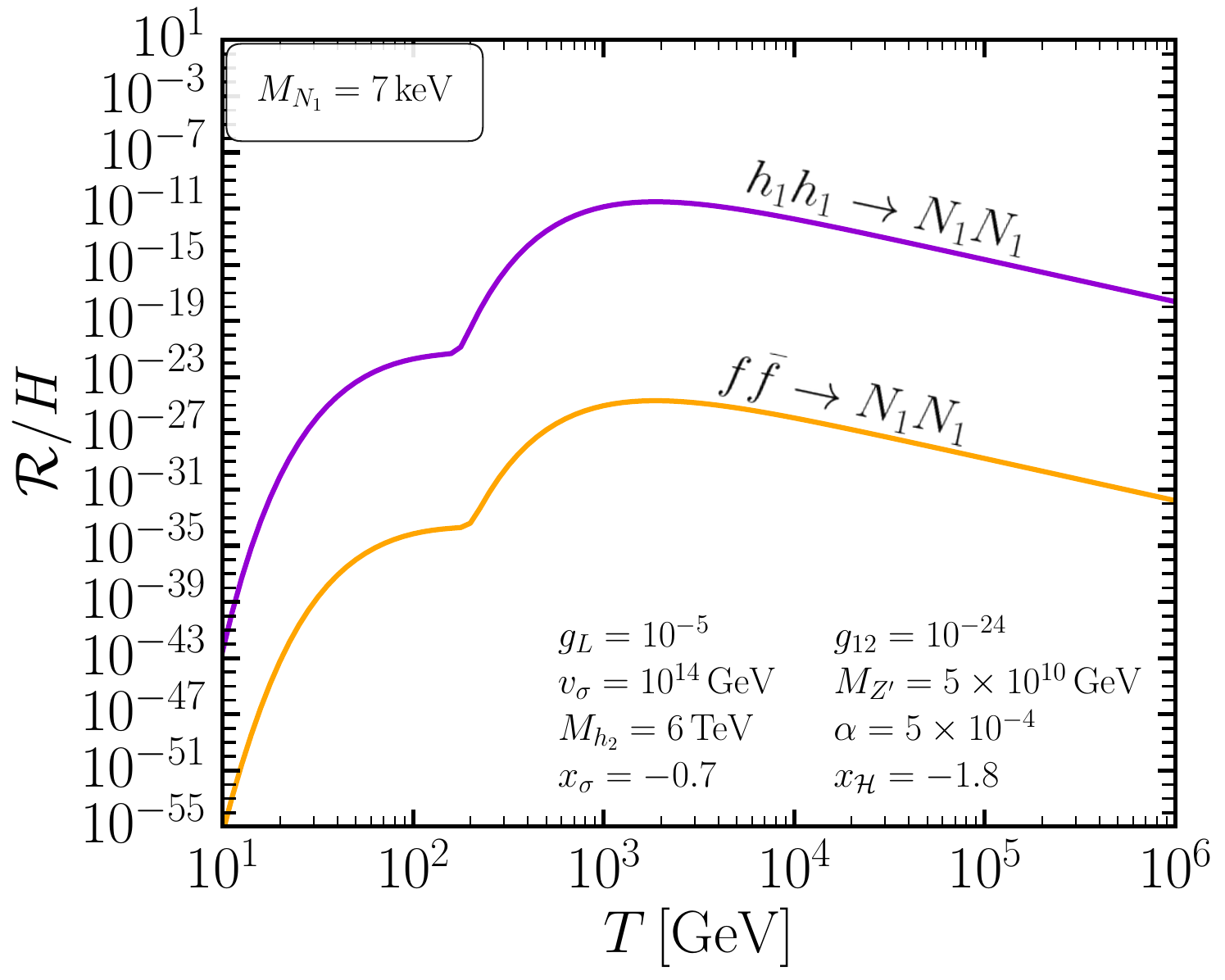}
  \end{minipage}
  \hfill
  \begin{minipage}{0.48\textwidth}
    \centering
    \includegraphics[width=\linewidth]{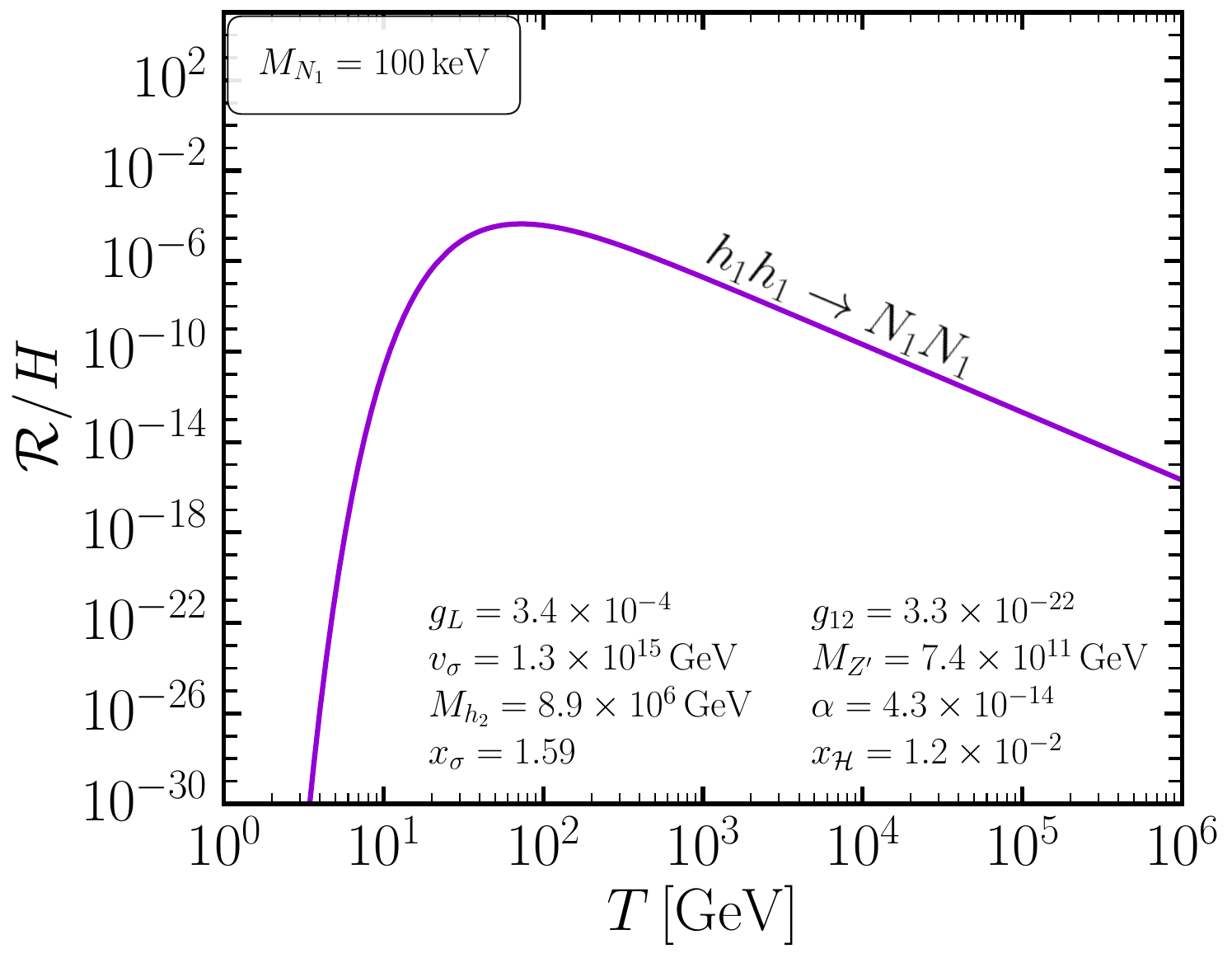}
  \end{minipage}
  \caption{Production  rate of $N_1$ divided by the Hubble expansion rate, $\mathcal{R}(\text{SM\;SM} \to N_I N_I)/H$, as a function of  temperature $T$ for two benchmark scenarios. The left panel corresponds to $M_{N_1}=7\,\mathrm{keV}$, where annihilation into fermionic final states is not very suppressed due to a relatively large value of the scalar mixing parameter $\alpha$. The right panel corresponds to $M_{N_1}=100$~keV, for which the Higgs channel dominates, and the other production channels are strongly suppressed because $\alpha$ is tiny. 
  }
  \label{fig:DM_Interaction_Rate}
\end{figure}

\begin{table}[t]
\centering
\begin{tabular}{l l l}
\hline
\textbf{Parameter} & \textbf{Description} & \textbf{Range } \\
\hline
$M_{h_2}$ & Heavy scalar mass & $150~\text{GeV} \lesssim M_{h_2} \lesssim 10^{7}~\text{GeV}  $ \\
$g_L$ & $\mathrm{U(1)^\prime}$ gauge coupling & $10^{-5} \lesssim g_L \lesssim 10^{-3}$ \\
$g_{12}$ & Kinetic mixing coupling & $10^{-25} \lesssim g_{12} \lesssim 10^{-15}$ \\
$x_\sigma$ & Scalar $\mathrm{U(1)^\prime}$ charge & $-5 < x_\sigma < 5$ \\
$x_\mathcal{H}$ & Higgs $\mathrm{U(1)^\prime}$ charge & $-5 < x_\mathcal{H} < 5$ \\
\hline
$\alpha$ & Scalar mixing angle & $10^{-15} \lesssim \alpha \lesssim 10^{-3}$ \\
$v_\sigma$ & Singlet VEV & $ 5 \times 10^{7} ~\text{GeV} \lesssim v_\sigma \lesssim 10^{16}~\text{GeV}$ \\
$M_{Z^\prime}$ & Dark gauge boson mass & $2 \times 10^5 ~\text{GeV} \lesssim M_{Z^\prime} \lesssim 3\times 10^{13}~\text{GeV}$ \\
\hline
\end{tabular}
\caption{Input parameter ranges in our numerical scan for models with nonthermal keV neutrino DM 
produced via freeze-in are shown in the upper part of the table. The output parameters in the bottom three rows are needed for the evaluation of the relic density. All parameters are evaluated at $\mu = 91~\mathrm{GeV}$. 
}
\label{tab:scan_limits}
\end{table}

To identify relevant freeze-in DM scenarios consistent with cosmological stability, nonthermalization, 
and experimental constraints, we perform wide scans over the parameter ranges  
in Table~\ref{tab:scan_limits}. All parameters are evaluated at the renormalization scale $\mu = 91$~GeV. We evolve the renormalization group equations~\cite{Goncalves:2024lrk} and verify that the model remains perturbative up to the grand unification scale. For each point we (i) impose Eq.~\eqref{eq:nontherm} together 
with cosmological stability of $N_1$, (ii) solve Eq.~\eqref{eq:Boltzmann_Y} 
for $Y_{N_1}^{\infty}$, and (iii) evaluate $\Omega_{N_1}h^2$ via Eq.~\eqref{eq:OmegaY}. While scenarios beyond 
the considered parameter ranges may feature nonthermal DM, they are expected to be less relevant
phenomenologically. 
For example, $g_L \lesssim 10^{-5}$
suppresses the interaction strength between $N_I$ and the SM bath, which in turn reduces 
the freeze-in efficiency yielding a too small relic density.

\begin{figure}[t]
   \centering
  \begin{minipage}{0.7\textwidth}
    \centering
    \includegraphics[width=\linewidth]{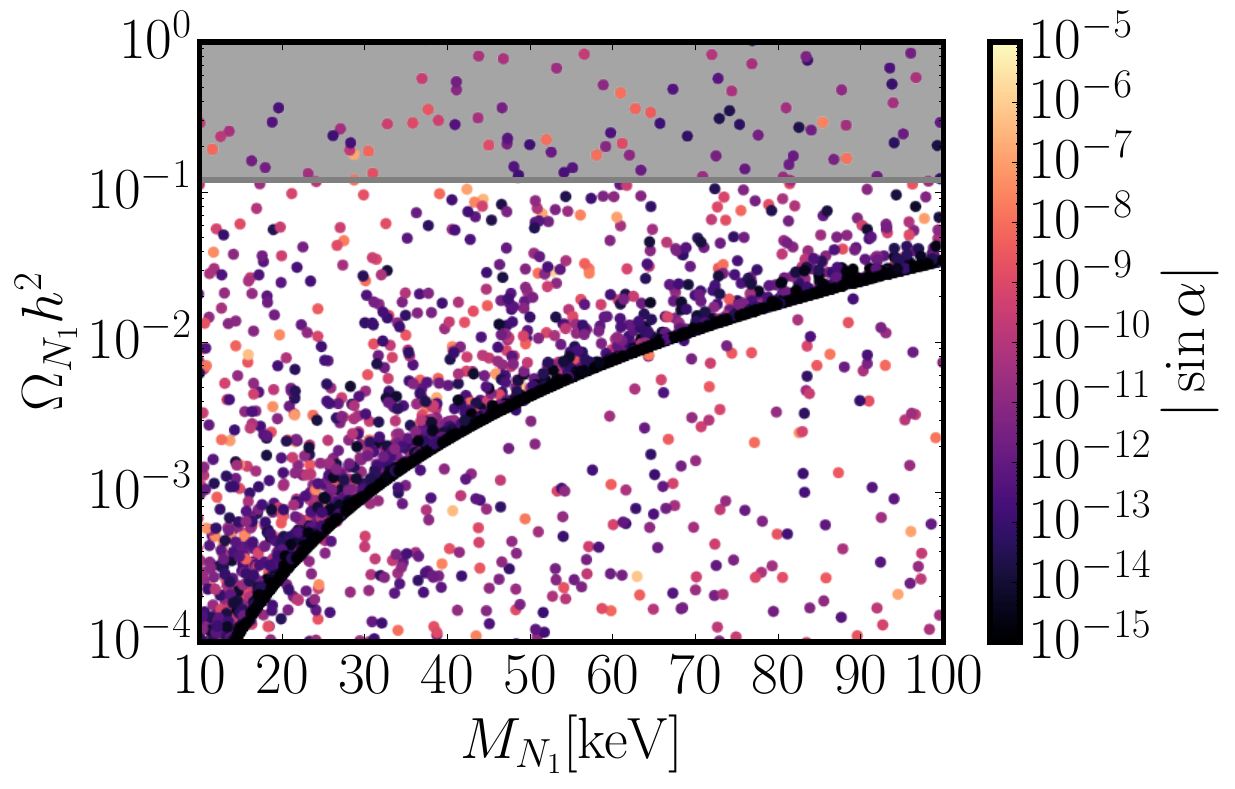}
  \end{minipage}
  \caption{$N_1$ relic abundance $\Omega_{N_1} h^2$ as a function of the DM  mass $M_{N_1}$ for different values of the scalar mixing parameter, $|\sin\alpha|$. All points satisfy the X-ray constraints.  The observed relic abundance $\Omega_{N_1} h^2 = 0.12$ is shown as a reference by gray line. Each point corresponds to the numerical solution of the Boltzmann equation (Eq.~\ref{eq:Boltzmann_Y}) accounting for all relevant $\mathrm{SM} \, \mathrm{SM}\to N_1 N_1$ production channels.
  }
  \label{fig:abundance}
\end{figure}

By numerically solving Eq.~\eqref{eq:Boltzmann_Y} for a strong mass hierarchy 
$M_{N_{2,3}} \gg M_{N_1}$, we obtain the abundance of the lightest sterile neutrino DM $N_1$ 
in Fig.~\ref{fig:abundance}. All parameter points displayed are consistent with the existing
constraints from X-ray observations. The DM relic density, $\Omega_{\rm DM} h^2 = 0.120 \pm
0.001$, inferred from \textit{Planck}~\cite{Planck:2018vyg}, is indicated by the dark grey horizontal
line. As mentioned earlier, our model favors small values of the scalar mixing angle $\alpha < 10^{-5}$.

Note that the $N_I$ production cross sections from SM fermion pairs, provided in Appendix~\ref{App:B}, are strongly suppressed by the heavy mediator $Z^\prime$  and $g_L\ll 1$. Similarly, the annihilation cross sections to SM vector bosons are suppressed by the mixing angle  $\alpha$. This indicates that the dominant $N_I$ production channels is via $h_1 h_1 \to N_I N_I$, while resonant enhancement occurs when $\sqrt{s}\simeq M_{h_2}$, leading to  a localized band of increased yield. In addition, the relevant $N_1$ couplings entering the production amplitudes  are proportional to $M_{N_1}$ in the seesaw/Majoron-like framework, which explains the $M_{N_1}$-dependence of the relic abundance. Consequently, for $\alpha \lesssim 10^{-14}$, the thermally averaged  cross section can be approximated by
 \begin{equation}\label{eq:tavg_xsec}
(n_{h_1}^{\text{eq}})^2 \langle \sigma(h_1h_1 \rightarrow N_1 N_1) v_r \rangle \simeq \frac{2\pi^2 }{(2\pi)^6} \frac{M_{h_1}^4 M_{N_1}^2 T^2}{16 \pi v_\sigma^4}[K_1(M_{h_1}/T)]^2\,,
\end{equation}
and the resulting freeze-in DM abundance takes the simplified form,
\begin{equation}\label{eq:approx}
\Omega_{N_1} h^2 \simeq 3\times 10^{-5} \,  \, \left( \frac{M_{N_1}}{10\; \text{keV}}\right)^3 \,.
\end{equation}
Remarkably, this cubic scaling with $M_{N_1}$ is clearly visible in Fig.~\ref{fig:abundance}.

Note that constraints imposed by X-ray searches can be relaxed by allowing for smaller active neutrino masses, which decreases the active-sterile mixing angle and weakens the bounds. In Fig.~\ref{fig:abundance}, the lightest active neutrino takes masses in the range $m_{\nu_1} \in [10^{-12},\,10^{-9}]\,\mathrm{eV}$ to ensure consistency with X-ray bounds.

\section{Lyman-\texorpdfstring{$\alpha$}{alpha} and strong gravitational lensing constraints}
\label{sec:LymanAlpha}

The freeze-in production mechanism generates $N_1$ with a nonthermal momentum distribution. Since primordial velocities suppress clustering below the corresponding 
freestreaming scale, such scenarios are tightly constrained by the Lyman-$\alpha$ forest, i.e., the absorption
pattern in distant quasar spectra produced by neutral hydrogen in the intergalactic medium (IGM). 
These measurements probe the matter power spectrum down to comoving wavenumbers 
$k\simeq[0.5\text{--}20]\,h/\mathrm{Mpc}$ and therefore provide some of the most stringent tests 
of non-cold DM whose primordial velocity distributions suppress structure formation on small scales.
Strong gravitational lensing provides a tool complementary to Lyman-$\alpha$ observations by measuring the halo mass function.
Non-cold DM particles have non-negligible velocities that suppress the formation of low-mass halos below the free-streaming scale. As a result, low-mass halos are both less abundant and less concentrated than for CDM, leading to observable signatures in strong lensing systems.

Current Lyman-$\alpha$ analyses are commonly expressed as lower limits on the mass of a \emph{thermal} 
warm-DM (WDM) relic. These limits depend on the assumed reionization history and on how the thermal evolution of the IGM is modeled.
 The most stringent limit, obtained from the combined XQ-100 
and HIRES/MIKE samples with standard IGM thermal priors~\cite{Irsic:2017ixq,Garzilli:2019qki}, requires
$m_{\mathrm{WDM}}\gtrsim 5.3~\mathrm{keV}$ (95\% C.L.) if WDM saturates the relic abundance. Allowing for non-monotonic 
temperature evolution of the IGM relaxes the bound to $m_{\mathrm{WDM}}\gtrsim
3.5~\mathrm{keV}$~\cite{Irsic:2017ixq}, while adopting colder reionization histories that minimize 
astrophysical pressure effects weaken it further to $m_{\mathrm{WDM}}\gtrsim
1.9~\mathrm{keV}$~\cite{Garzilli:2019qki}. 
On the other hand, recent analyses of strongly lensed quasars observed with JWST  significantly strengthen the bound to $m_{\rm WDM} \gtrsim 8.4~\mathrm{keV}$ (95\% C.L.)~\cite{Gilman:2025fhy}.
 We therefore consider three benchmark limits, $m_{\rm WDM}=8.4~\mathrm{keV}$, $5.3~\mathrm{keV}$, and $1.9~\mathrm{keV}$, which span the strongest current JWST strong-lensing benchmark and the stringent/conservative Lyman-$\alpha$ benchmarks. Note that the strong-lensing and Lyman-$\alpha$ data are subject to completely different systematic uncertainties. 

To confront our nonthermal $N_1$ with these limits we compute the linear matter power spectrum $P(k)$ 
using \texttt{CLASS}~\cite{blas2011cosmic}.  
 The $N_1$ momentum distribution resulting 
from freeze-in is implemented in \texttt{CLASS} as a non-cold DM species. For DM produced through $2\to 2$
scattering, the phase-space distribution is well approximated by the analytical form~\cite{DEramo:2020gpr},
\begin{equation}
f(q)\propto \frac{e^{-q}}{\sqrt{q}}\,,
\label{eq:fq_analytic}
\end{equation}
where $q\equiv p/T$.
We have explicitly verified by solving the Boltzmann equation that 
the freeze-in distribution of $N_1$ follows Eq.~\eqref{eq:fq_analytic} even at resonance. A  derivation of the phase-space distribution in the resonance regime is presented in Appendix~\ref{app:fq-resonance}.
The overall normalization is fixed for each point in parameter space such that the resulting energy density reproduces the required 
$N_1$ abundance. 

To quantify the suppression of small-scale power relative to $\Lambda$CDM (the {\it $\Lambda$ cold dark matter} cosmology with Planck's 2018 best-fit parameters~\cite{Planck:2018vyg}) and recast it 
into thermal-WDM bounds, we adopt the \emph{half-mode} criterion~\cite{Schneider:2016uqi,Murgia:2017lwo,DEramo:2020gpr}, which provides a mapping between nonthermal relics and thermal WDM relics constrained  by the Lyman-$\alpha$ forest and  strong lensing~\cite{Zelko:2022tgf}. We first define the squared transfer function,
\begin{equation}
\mathcal{T}^2(k)\equiv \frac{P(k)}{P_{\Lambda\mathrm{CDM}}(k)}\,,
\label{eq:transfer}
\end{equation}
which encodes the small-scale power suppression of the nonthermal relic with respect to $\Lambda$CDM. We then characterize the cut-off scale for power suppression through the \emph{half-mode} wavenumber $k_{1/2}$, defined as the scale at which the squared transfer function drops by a factor of two,
\begin{equation}
\mathcal{T}^2(k_{1/2})\equiv \frac{1}{2}\,.
\label{eq:halfmode_def}
\end{equation}

We compute $k_{1/2}$ for nonthermal $N_1$ and compare it to the reference values $k_{1/2}^{\mathrm{WDM}}$ obtained for thermal WDM relics at the benchmark limits, $m_{\mathrm{WDM}}=8.4~\mathrm{keV}$, $m_{\mathrm{WDM}}=5.3~\mathrm{keV}$ and $m_{\mathrm{WDM}}=1.9~\mathrm{keV}$, and exclude parameter points for which
\begin{equation}
k_{1/2}<k_{1/2}^{\mathrm{WDM}}\,.
\label{eq:halfmode_crit}
\end{equation}
Physically, a half-mode below the WDM reference means that the small-scale cut-off of the $N_1$ power spectrum occurs at scales that are too large (wavenumbers too small) to be compatible with the data. Using \texttt{CLASS} to compute the thermal WDM transfer functions, we find 
\begin{equation}
k_{1/2}^{\mathrm{WDM}}\simeq
\begin{cases}
92.1~h/\mathrm{Mpc} & m_{\mathrm{WDM}}=8.4~\mathrm{keV}\\[2pt]
52.5~h/\mathrm{Mpc} & m_{\mathrm{WDM}}=5.3~\mathrm{keV}\\[2pt]
15.6~h/\mathrm{Mpc} & m_{\mathrm{WDM}}=1.9~\mathrm{keV}\,.
\end{cases}
\label{eq:halfmode_WDM}
\end{equation}
\begin{figure}[t]
\centering
\includegraphics[width=0.6\textwidth]{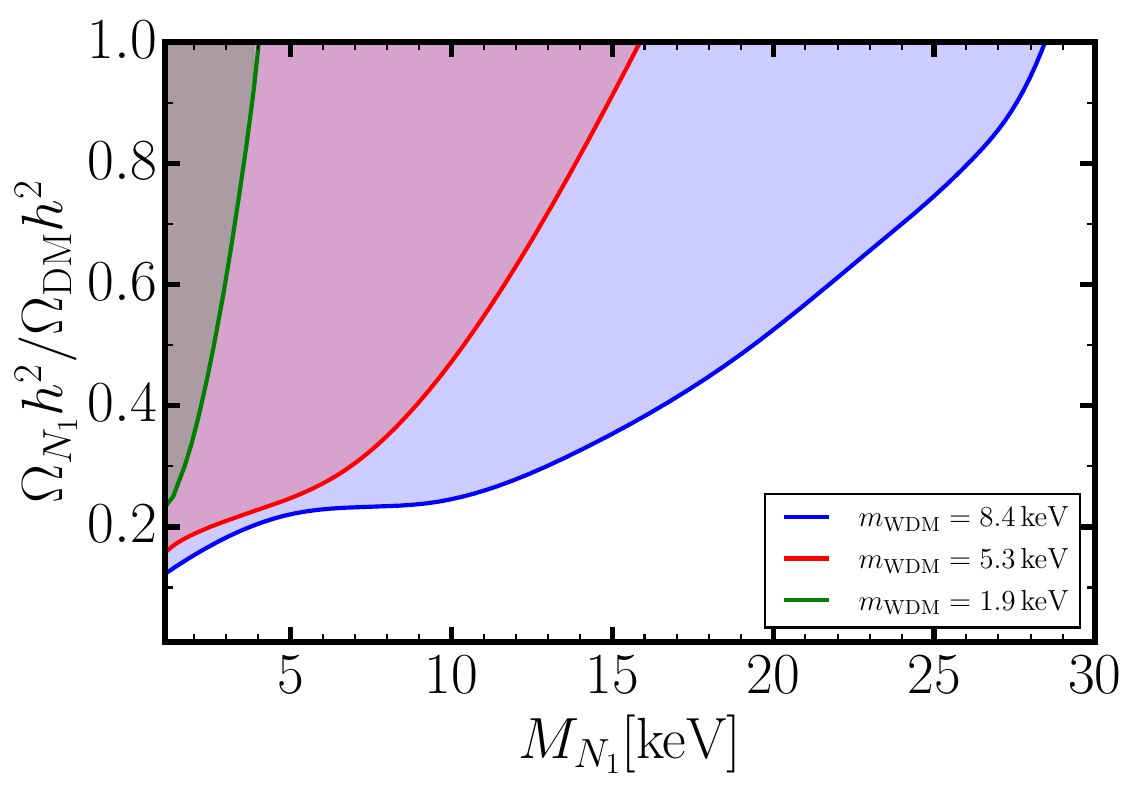}
\caption{Fraction of the DM abundance contributed by $N_1$ as a function of $M_{N_1}$. The curves show the 95\%~C.L. upper limits derived from Lyman-$\alpha$ forest and strong-lensing constraints, using the equivalent thermal WDM masses $m_{\rm WDM} = 8.4\,\mathrm{keV}$ (blue),
$m_{\rm WDM} = 5.3\,\mathrm{keV}$ (red) and $m_{\rm WDM} = 1.9\,\mathrm{keV}$ (green). 
}
\label{fig:lyman}
\end{figure}

Figure~\ref{fig:lyman} shows $\Omega_{N_1} h^2 / \Omega_{\rm DM} h^2$ as a function of $M_{N_1}$. The red and green curves correspond to the 95\%~C.L. Lyman-$\alpha$  upper limits  recast from the stringent and conservative thermal WDM limits, respectively. The blue curve corresponds to the 95\%~C.L. upper limit from JWST data. 
 In the limit that $N_1$ constitutes all of DM,  the stringent Lyman-$\alpha$ bound requires $M_{N_1}\gtrsim 15.8~\mathrm{keV}$, whereas the conservative bound requires $M_{N_1}\gtrsim
4.0~\mathrm{keV}$. The JWST data constrain $M_{N_1} \gtrsim 28.3~\mathrm{keV}$. 

\section{$3.5$~keV X-ray line}
\label{sec:35keVline}

A particularly intriguing phenomenological target for keV sterile neutrino DM is the
weak  X-ray feature at $E_\gamma\simeq 3.52\pm0.02~\mathrm{keV}$ in stacked
observations of galaxy clusters and individual systems such as the Perseus cluster and 
M31, with a global significance of about $4.4\sigma$~\cite{Bulbul:2014sua,Boyarsky:2014jta}.
Despite the uncertain status of the $3.5~\mathrm{keV}$ line, we take an optimistic approach and consider the historically reported $3.5~\mathrm{keV}$ line interpretation as a phenomenological benchmark.

The line can be interpreted as the monoenergetic photon arising from the two-body radiative transition $N_1\to \nu+\gamma$. The dominant contribution to the decay rate
is given by Eq.~\eqref{eq:rad_gammadecay}. The inferred 
signal region of the parameter space,
\begin{equation}
M_{N_1}\simeq 7~\mathrm{keV}\,,\qquad
\sin^2 2\theta_{\rm eff}\sim (0.2{-}2)\times10^{-10}\,,
\label{eq:35keVtarget}
\end{equation}
corresponds to a cosmologically long-lived particle~\cite{Drewes:2016upu}.  We assess whether our Higgs-dominated freeze-in scenario 
can simultaneously reproduce the inferred DM relic abundance and the mixing range needed.
\begin{figure}[t]
    \centering
    \includegraphics[width=0.7\textwidth]{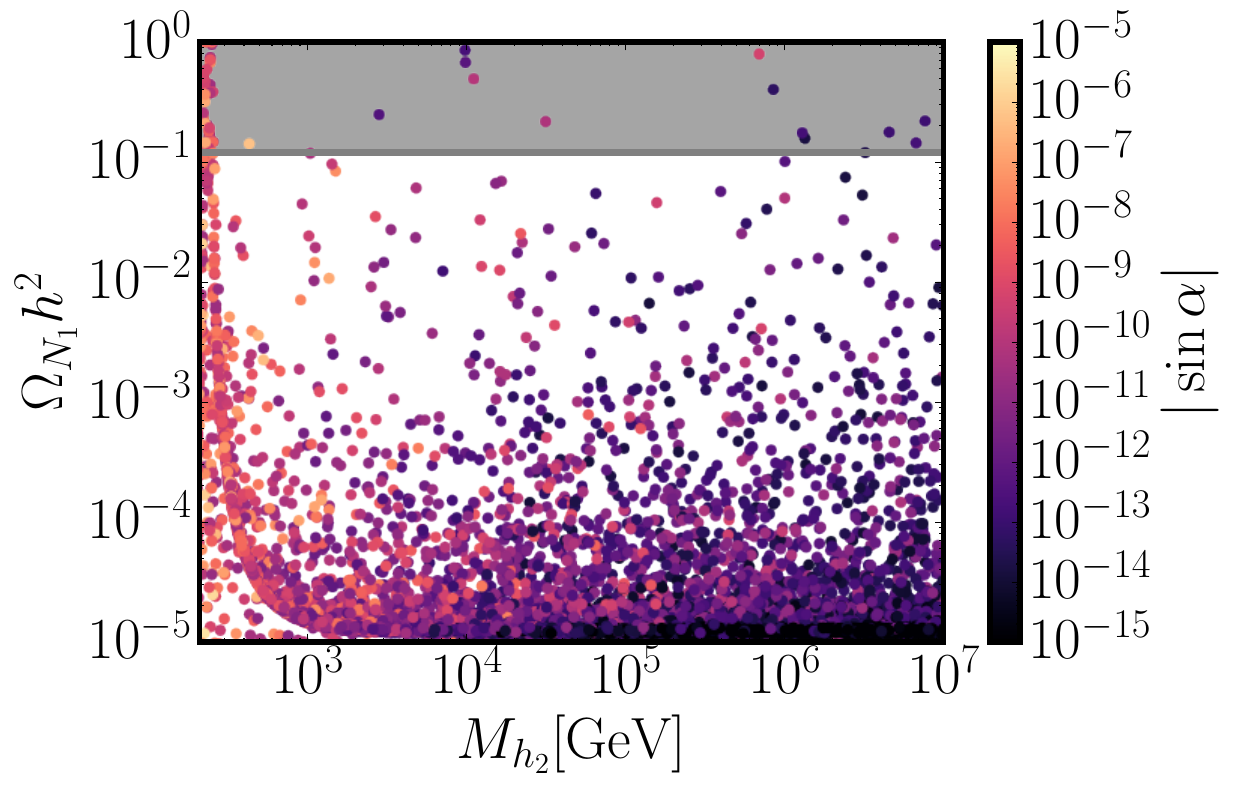}
     \caption{$N_1$ relic abundance for $M_{N_1}=7~\mathrm{keV}$. The color scale shows the corresponding values of the scalar mixing $|\sin\alpha|$. The gray line indicates the inferred DM abundance, $\Omega_{\rm DM}h^2\simeq 0.12$. Each
     point is obtained by solving the Boltzmann equation (Eq.~\ref{eq:Boltzmann_Y}) with the thermally averaged
     annihilation rate including all relevant $\mathrm{SM} \, \mathrm{SM}\to N_1 N_1$ production channels.
  }
    \label{fig:Abundance_7keV}
\end{figure}

The present-day DM abundance for a 7~keV neutrino is shown in Fig.~\ref{fig:Abundance_7keV} as a function of $M_{h_2}$, where 
the color scale indicates the value of $|\sin\alpha|$. Two distinct regimes can be identified. 

For $M_{h_2}\lesssim \mathcal{O}(10^3)$~GeV, relatively larger mixings $|\sin\alpha|\sim
\mathcal{O}(10^{-6}{-}10^{-5})$ and smaller VEVs 
$v_\sigma\sim\mathcal{O}(10^{10}{-}10^{12})$~GeV, are required. This combination does not greatly suppress the interaction between  $N_1$ and the SM bath and therefore enhances the freeze-in production rate. More importantly, for smaller $M_{h_2}$, $N_1$ production is resonantly enhanced in the Higgs channel. 
For the dominant process  $h_1h_1\to N_1N_1$ the threshold $s_{\min}=4M_{h_1}^2$, and if $M_{h_2} \lesssim 2M_{h_1}$ is close to the Higgs-pair threshold, then the thermally averaged cross section and freeze-in yield are enhanced.
A small enhancement in production through SM fermions can also arise in this region: lowering $M_{h_2}$
correlates with parameter choices that reduce the effective suppression of the $Z^\prime$-mediated
contribution, thereby increasing the rate for $f\bar f\to N_1N_1$. However, we find that this 
contribution remains subdominant with respect to Higgs-initiated production in the parameter space 
relevant to our $7$~keV benchmark, although it contributes to the shape of the region at small values of $M_{h_2}$.

For $M_{h_2}\gtrsim \mathcal{O}(10^3)$~GeV, $v_\sigma$ increases and $|\sin\alpha|$ decreases, suppressing the scalar-mediated contribution. Also, the resonance condition $s\simeq M_{h_2}^2$ moves into the large-$s$ tail of the thermal integral, where the Bessel function $K_1(\sqrt{s}/T)$ exponentially suppresses the integrand. As a result, the $N_1$ abundance rapidly decreases with increasing $M_{h_2}$ and approaches the asymptotic Higgs-dominated regime for large $h_2$ masses. The viable points then accumulate around $\Omega_{N_1} h^2\sim 10^{-5}$ for large $M_{h_2}$, consistent with the scaling in Eq.~\eqref{eq:approx}.
\begin{figure}[t!]
    \centering
    \includegraphics[width=0.7\textwidth]{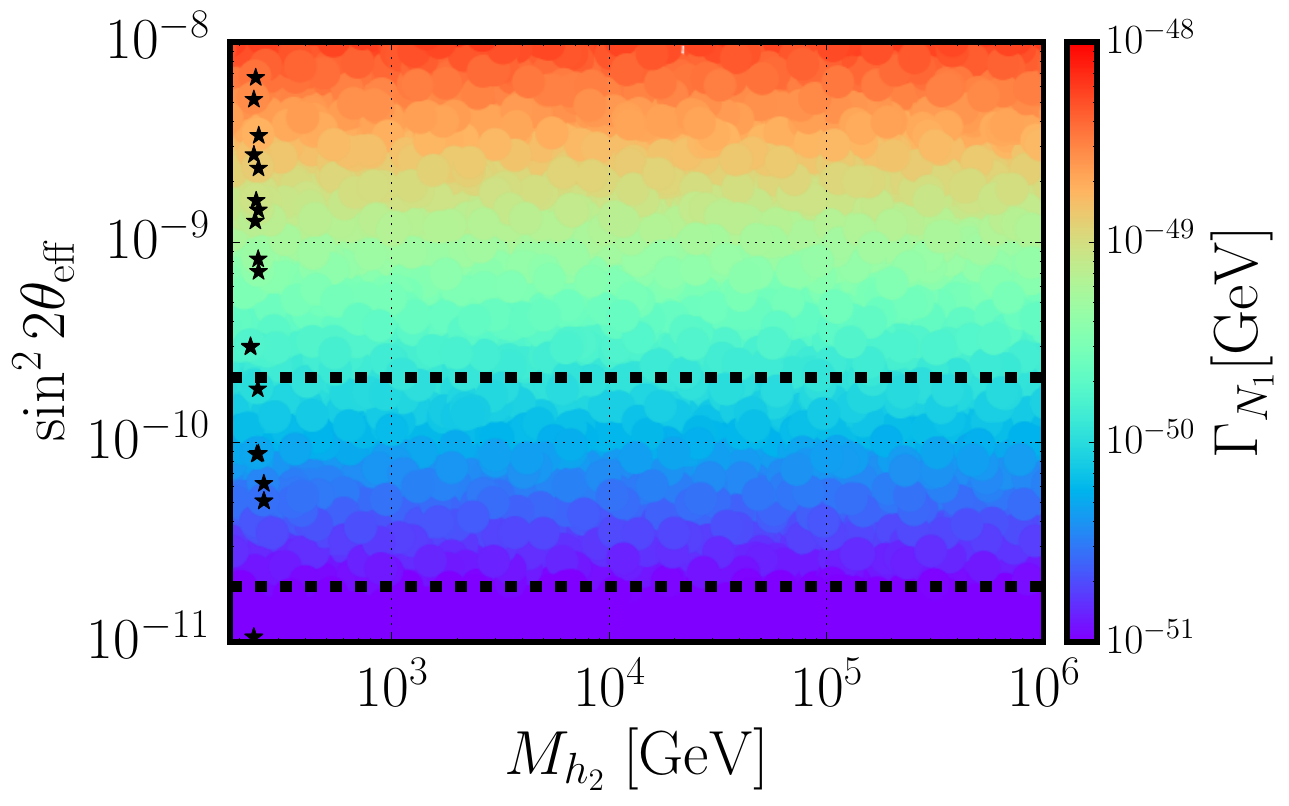}
    \caption{$\sin^2 2\theta_{\rm eff}$ as a function of $M_{h_2}$ for parameter
    points satisfying $\Omega_{N_1}h^2\le 0.12$. The color scale indicates $\Gamma_{N_1}$, and the black 
    stars mark points with $\Omega_{N_1}h^2=0.12\pm0.001$. The region between
    dashed lines corresponds to the range $\sin^2 2\theta_{\rm eff}\simeq (0.2{-}2)\times10^{-10}$ needed to produce a $3.5$~keV X-ray line via $N_1\to\nu\gamma$. 
    }
    \label{fig:sin2thetaeff_abundance}
\end{figure}

In Fig.~\ref{fig:sin2thetaeff_abundance} we show  
$\sin^2 2\theta_{\rm eff}$ as a function of $M_{h_2}$ for points satisfying $\Omega_{N_1}h^2\le 0.12$.
The color scale indicates the value of $\Gamma_{N_1}$, and the black stars mark the points
with $\Omega_{N_1} h^2=0.12\pm0.001$. The region between the dashed 
lines corresponds to the mixing angles needed to reproduce the X-ray line at $E_\gamma\simeq
3.5~\mathrm{keV}$. The stars accumulate at the Higgs threshold discussed above. Remarkably, some of them fall within the X-ray band. One of these has the following parameters:
\begin{align} \nonumber
&M_{Z^\prime} =  10^{7}~\mathrm{GeV}\,,\quad
M_{h_2} = 244 ~\mathrm{GeV}\,,\quad
g_L = 5.5\times10^{-5}\,,\quad
\alpha = 9.0\times10^{-10}\,,\\
&v_\sigma =  10^{11}~\mathrm{GeV} \,,\quad
x_\mathcal{H} = - 1.6\times10^{-5} \,,\quad
x_\sigma = 2 \,,\quad g_{12} = 10^{-23}\,,
\label{eq:benchmark-7keV}
\end{align}
and provides a natural benchmark for the $7~\mathrm{keV}$ sterile neutrino scenario that
simultaneously reproduces the observed relic abundance and the mixing range relevant 
for the $3.5$~keV line. 
Note that this point is allowed only by the conservative Lyman-$\alpha$ limit in Fig.~\ref{fig:lyman}.

\section{Multi-component decaying dark matter and the $S_8$ tension}
\label{sec:S8}

A well-motivated mechanism capable of reducing the predicted clustering amplitude 
at intermediate scales is \emph{decaying dark matter} (DDM), in which a long-lived parent
decays on cosmological time scales to a slightly lighter daughter  that receives 
a non-relativistic momentum ``kick'', thereby suppressing the growth of
structure on small scales~\cite{Fuss:2024dam,Astros:2025aan}. Recent DESI-DR1 joint analyses with weak lensing datasets 
find values of $S_8$ at the level $S_8\simeq 0.76$-$0.79$ (depending on dataset combinations and
priors)~\cite{Porredon:2025xxv,Semenaite:2025ohg,Lange:2025rsx}, consistent with the range for 
which DDM-like suppression mechanisms are relevant.

In our framework, this scenario can be realized in a minimal way if at least two keV-scale sterile 
neutrinos are long-lived. We identify $N_2$ as the parent state and $N_1$ as the daughter state. 
The suppression of the matter power spectrum is controlled primarily by the parent lifetime 
$\tau_{N_2}$ and by the fractional mass splitting, which we parameterize as~\cite{Fuss:2024dam},
\begin{equation}
\epsilon \equiv \frac{1}{2}\left(1-\frac{M_{N_1}^2}{M_{N_2}^2}\right)
\;\simeq\;\frac{M_{N_2}-M_{N_1}}{M_{N_2}}\ll 1 \,, \qquad
M_{N_2}\gtrsim M_{N_1}\,.
\label{eq:eps_def}
\end{equation}
Physically, $\epsilon$ controls the kinetic energy imparted to the daughter 
and therefore the amount of freestreaming suppression, while $\tau_{N_2}$ determines the freestreaming wavenumber.

Following Ref.~\cite{Fuss:2024dam}, we identify a representative region that matches the lifetime and mass-splitting regime known to alleviate $S_8$ tension:
\begin{equation}
\tau_{N_2}\sim (4{-}8)\times 10^{18}~\mathrm{s} \,,
\qquad
\epsilon \sim 0.01{-}0.1\,,
\label{eq:S8window}
\end{equation}
where the precise preferred window depends on the late-time datasets 
and on the mapping from the power-spectrum suppression to the inferred $S_8$. 
Since the recent results by DESI~\cite{Porredon:2025xxv,Semenaite:2025ohg,Lange:2025rsx} 
continue to prefer $S_8$ values in the range for which DDM remains a viable phenomenological 
option, we adopt Eq.~\eqref{eq:S8window} as an indicative target region and test whether it can be realized in our model.
\begin{figure}[t!]
    \centering
    \includegraphics[width=0.495\textwidth]{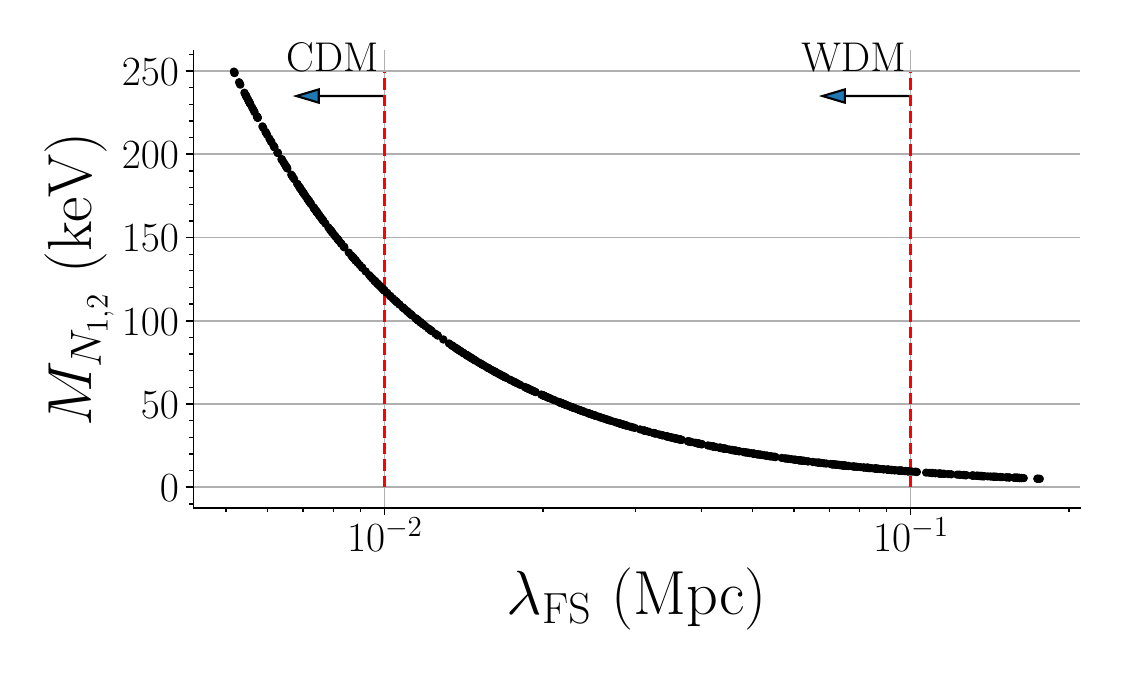}
    \includegraphics[width=0.495\textwidth]{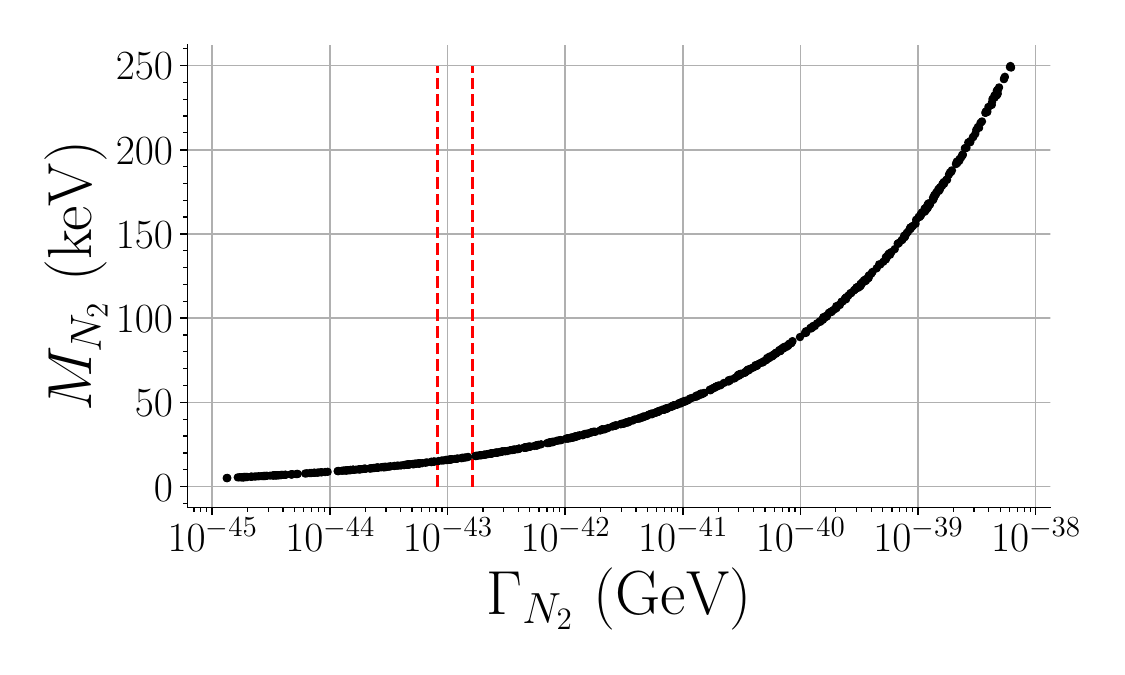}
    \caption{
    Left: Freestreaming length $\lambda_{\rm FS}$ of the quasi-degenerate neutrinos $N_1$ and $N_2$. The vertical dashed 
    lines indicate the regions where the sterile neutrinos behave as either cold or warm DM.
    Right: Mass of the parent sterile neutrino $N_2$ versus its total decay width $\Gamma_{N_2}$; 
    the vertical dashed lines indicate the decay-width range corresponding to 
    $\tau_{N_2}\sim (4{-}8)\times10^{18}\,\mathrm{s}$. }
    \label{fig:MN5_gamma_lambda}
\end{figure}

Unlike the minimal DDM setups studied in~\cite{Fuss:2024dam,Astros:2025aan}, the daughter state $N_1$ is not stable 
in our model. Therefore, to determine whether a viable hierarchy of lifetimes can be achieved, we compute all 
relevant two- and three-body decay channels for both $N_2$ and $N_1$, including $N_I\to \nu\gamma$, 
$N_I\to \nu\bar\nu\nu$, and $N_2\to N_1\bar\nu\nu$.
In addition, while the canonical DDM picture assumes a cold parent component, this need not be the case.
In the parameter region relevant to Eq.~\eqref{eq:S8window} both $N_1$ and $N_2$ can behave as warm relics. 
To quantify this, we compute the freestreaming length $\lambda_{\rm FS}$ following~\cite{DEramo:2020gpr} and present 
the result in Fig.~\ref{fig:MN5_gamma_lambda} (left). We find that the parent is cold only 
for $M_{N_2}\gtrsim 120~\mathrm{keV}$. Comparing with the decay-width window in
Fig.~\ref{fig:MN5_gamma_lambda} (right), the region compatible with Eq.~\eqref{eq:S8window} corresponds to a narrow 
mass interval $M_{N_2}\simeq 15~\mathrm{keV}-20~\mathrm{keV}$, which lies in the warm regime. 

In this narrow mass range we find that the model can realize a pronounced hierarchy between
$\tau_{N_2}$ and $\tau_{N_1}$ as shown in the left panel of Fig.~\ref{fig:lifetime}. $N_2$ decays on cosmological
time scales specified by Eq.~\eqref{eq:S8window}, whereas $N_1$ remains much longer-lived and behaves 
as a quasi-stable warm particle. In this sense, $N_2$ plays the role of a decaying warm DM component while 
$N_1$ constitutes the effectively stable warm component.
\begin{figure}[t!]
    \centering
    \includegraphics[width=\textwidth]{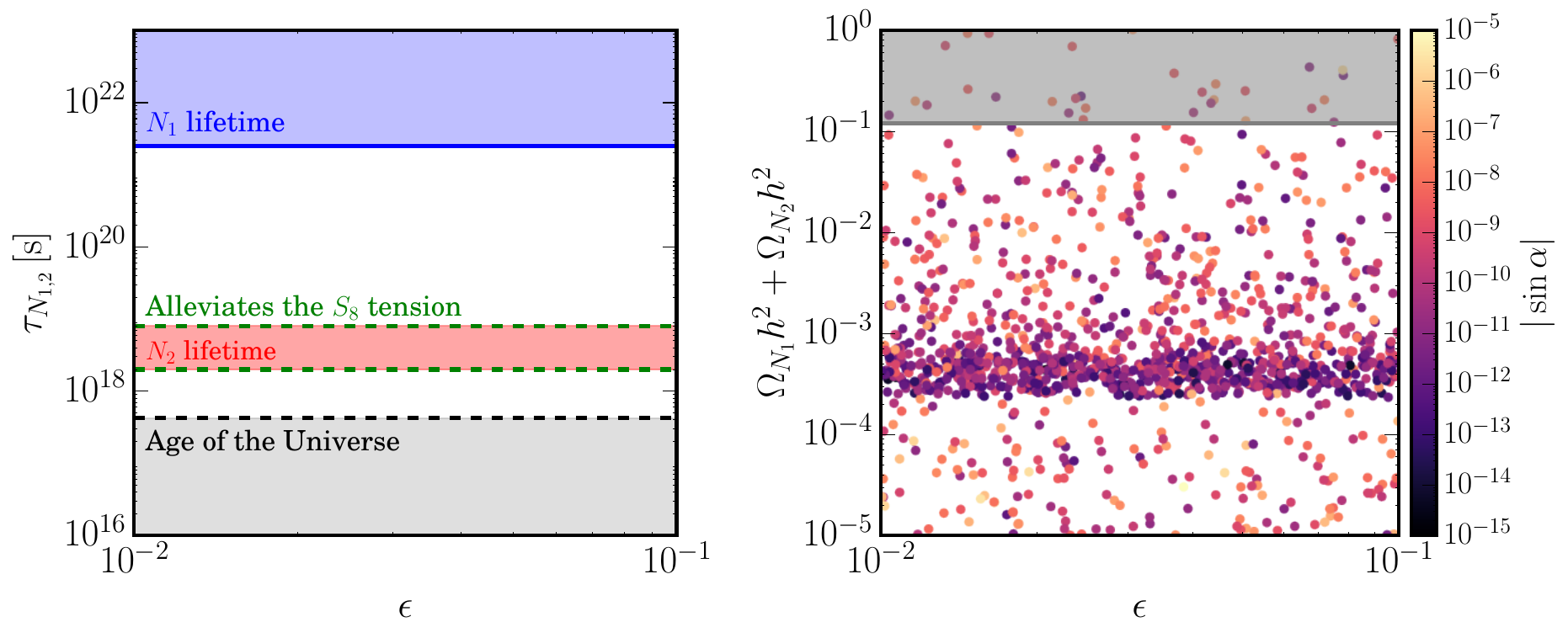}
    \caption{
    Left: Lifetimes of $N_1$ and $N_2$ as functions of the mass-splitting parameter $\epsilon$;  the shaded band highlights the  range $\tau_{N_2}\sim (4{-}8)\times10^{18}\,\mathrm{s}$ that alleviates the $S_8$ tension. Right: Relic abundance of the sterile neutrinos for different values of the scalar mixing $|\sin\alpha|$. 
}
    \label{fig:lifetime}
\end{figure}

Note that the freeze-in picture developed in Section~\ref{sec:freezein} continues to apply.
Since $N_1$ and $N_2$ are quasi-degenerate and are generated through the same suppressed portals, the freeze-in yields 
are comparable, and the total sterile neutrino abundance described by Eq.~\eqref{eq:Boltzmann_Y} 
can be approximated as $Y \equiv Y_{N_1}+Y_{N_2}\simeq 2\,Y_{N_2}$, in analogy to the decaying CDM setup of 
Ref.~\cite{Fuss:2024dam}. Consequently, the total relic density is naturally shared between the two species, 
with each contributing an $\mathcal{O}(1)$ fraction.  Importantly, in the region of parameter space satisfying
Eq.~\eqref{eq:S8window}, we find that the correct total relic abundance can be achieved, as illustrated 
in Fig.~\ref{fig:lifetime} (right), while maintaining the required lifetime hierarchy.
Although $ M_{N_1}\simeq M_{N_2} \approx 15~\mathrm{keV}-20~\mathrm{keV}$ is not excluded by Lyman-$\alpha$ data, it is incompatible with the strong-lensing limit $M_{N_1}\gtrsim28~\mathrm{keV}$; see Fig.~\ref{fig:lyman}.

To be clear, a fully consistent assessment of the $S_8$ tension in our framework entails evolving the phase-space distributions of $N_2$ and $N_1$, and computing the linear matter power spectrum at the $\sigma_8$ and Lyman-$\alpha$/strong-lensing scales.
Since both neutrinos are produced non-thermally via freeze-in, an implementation in  \texttt{CLASS} involves a substantial modification of the source code, and is beyond the scope of this work.

\section{The KM3NeT event}
\label{sec:EeV-scale}

Before concluding, we briefly note that the conformal $\mathrm{U(1)^\prime}$ framework can also accommodate much 
heavier sterile neutrino DM candidates, in a different corner of parameter space. This possibility has attracted 
attention following the KM3NeT event, KM3-230213A, a track-like signature compatible with 
an ultra-high-energy muon of reconstructed energy $E_\mu = 120^{+110}_{-60}~\mathrm{PeV}$~\cite{KM3NeT:2025npi}. The parent neutrino energy is inferred to be in the broad range 
$E_\nu \sim 110~\mathrm{PeV}-790~\mathrm{PeV}$ (at $1\sigma$), with a median of
$220~\mathrm{PeV}$~\cite{KM3NeT:2025npi}. If the event is interpreted as originating 
from the two-body decay of a superheavy DM particle into light neutrinos, the characteristic mass scale is set by 
$M_{\rm DM}\sim 2E_\nu$, pointing to $M_{\rm DM}\sim 440~\mathrm{PeV}$.

We find that our model can reproduce such a heavy sterile neutrino. However, this solution requires an extreme 
amount of fine-tuning because the DM needs to be exceptionally long-lived,
which in a minimal type-I seesaw realization corresponds to ultra-suppressed active-sterile Yukawa couplings. Using $y_\nu \sim \mathcal{O}(10^{-31})$ for
$M_{N_1}\simeq 440~\mathrm{PeV}$, we find (in the absence of flavor cancellations) the scale of the lightest neutrino mass to be
\begin{equation}
m_\nu \;\sim\; \frac{y_\nu^2\,v^2}{2M_{N_1}}
\;\sim\;10^{-56}\ \mathrm{eV}\,,
\end{equation}
which illustrates the degree of tuning required.

In addition, unlike the freeze-in production mechanism discussed in Section~\ref{sec:freezein} for keV DM, the relic 
density for $M_{N_1} \simeq 440~\mathrm{PeV}$ cannot be efficiently generated through the same
$2\to2$ bath processes. In our model, $M_{Z^\prime}>M_{h_2}$, so that $h_2\to N_1\bar N_1$ is
the only efficient production mode with $h_2$ effectively acting as the reheating/inflaton field.
Obtaining the correct relic abundance then requires the decay to occur extremely close to threshold, with a 
relative mass splitting,
\begin{equation}
\frac{M_{h_2}-2M_{N_1}}{M_{h_2}}\;\sim\;10^{-12}\,,
\end{equation}
so that phase-space suppression compensates the otherwise excessive population of $N_1$ from $h_2$ decays.

\section{Summary}
\label{sec:conclusions}

We studied a classically conformal $\mathrm{U(1)^\prime}$ extension of the SM in which the lightest 
sterile neutrino, $N_1$, is a feebly interacting dark-matter (DM) candidate produced via freeze-in. The conformal 
symmetry forbids explicit mass scales in the tree-level Lagrangian, and the full spectrum, including 
the $\mathrm{U(1)^\prime}$ gauge boson $Z^\prime$ and the singlet-like scalar $h_2$, is generated dynamically 
via radiative corrections. In the minimal keV-DM setup, $N_1$ is the only long-lived sterile state. If the 
next-to-lightest sterile neutrino $N_2$ is also long-lived, the model naturally realizes multi-component 
DM produced via freeze-in with late-time decays.

We performed a numerical calculation of the sterile neutrino abundance by solving the Boltzmann equation for 
freeze-in production, scanning the parameter space and identifying regions consistent with the observed 
relic density, $\Omega_{\rm DM}h^2\simeq 0.12$. We implemented the corresponding nonthermal phase-space 
distribution in the Boltzmann solver \texttt{CLASS} and computed the linear matter power spectrum. 
We interpreted Lyman-$\alpha$ forest and strong lensing constraints through the thermal-equivalent WDM benchmarks, and found 
that the predicted suppression of small-scale power is 
compatible with these limits within the viable freeze-in parameter space.

In the keV mass range, the radiative decay $N_1\to\nu\gamma$ provides a clear X-ray line signal. In particular, for $M_{N_1}=7~\mathrm{keV}$ we identified a viable parameter region
that simultaneously reproduces the inferred DM abundance and the effective mixing angle relevant for 
interpreting the unidentified line at $E_\gamma\simeq 3.5~\mathrm{keV}$ observed in galaxy 
clusters as due to $N_1\to\nu\gamma$ decays.
This explanation survives the conservative Lyman-$\alpha$ limit, but not the stringent Lyman-$\alpha$ or strong-lensing limits in Fig.~\ref{fig:lyman}.

If both $N_1$ and $N_2$ are long-lived, the model realizes a decaying (multi-component) DM setup in which 
the late decay $N_2\to N_1+\mathrm{(light~states)}$ injects a small velocity dispersion for the daughter particle and can suppress the growth of structure on small scales. We demonstrated that the parameter space can accommodate the indicated lifetime and mass-splitting window relevant for alleviating the $S_8$ tension while reproducing the correct relic abundance. Although this explanation is compatible with Lyman-$\alpha$ data, it is incompatible with the strong-lensing limit from JWST.

Finally, we commented on a corner of parameter space in which the decay of a 440~PeV sterile neutrino 
$N_1$, being the lightest in the Majorana neutrino spectrum, can play the role of DM and explain the KM3NeT event, KM3-230213A. However, an exceptional degree 
of fine-tuning in the neutrino-sector parameters and in the nonthermal cosmological history (notably, 
near-threshold production from $h_2$ decays) is required. We, therefore, regard the keV-scale freeze-in scenario 
as the most robust and predictive regime of the model.

\section*{Acknowledgments}
D.M.~is supported in part by the U.S. Department of Energy under Grant No.~DE-SC0010504.
A.P.M.~is supported by FCT through the project with reference 2024.05617.CERN (\url{https://doi.org/10.54499/2024.05617.CERN}).
J.G.~and A.P.M.~are also supported by LIP and FCT, reference LA/P/0016/2020 (\url{https://doi.org/10.54499/LA/P/0016/2020}) and by the ERC-PT A-Projects 'Unveiling', financed by PRR, NextGenerationEU.
V.O. is directly funded by FCT through the doctoral program grant with the reference PRT/BD/154629/2022 (\url{https://doi.org/10.54499/PRT/BD/154629/2022}). V.O. also acknowledges support by the COST Action
CA21106 (Cosmic WISPers).

\appendix
\section{Production cross sections for light dark matter}
\label{App:B}

In the freeze-in regime, the sterile neutrino relic abundance is sourced by feeble $2\to2$ processes 
involving SM particles in the thermal bath. For analytic insight, we consider the parametric limit
\begin{equation}
M_{N_1}\ll M_{\rm SM},\qquad |\alpha|\ll 1,
\end{equation}
and retain only the leading contributions from $s$-channel exchange. The expressions below are 
order-of-magnitude approximations that reproduce the qualitative behavior and the numerical trends 
of the full computation in the relevant parameter space. Here, $s$ denotes 
the usual Mandelstam variable, and mediator widths in the propagators can be 
reinstated by the substitution, $M^2\to M^2-iM\Gamma$.

We list production cross sections in the limit that the kinematic dependence on $M_{N_1}$ is negligible 
and the overall coupling strength is determined by the feeble portals.
\begin{align}
\sigma_{\ell^+ \ell^- \to N_1 N_1 }
&\simeq
\frac{g_L^4\,x_\sigma^2}{384\pi}\,
\frac{s\bigl(10 x_{\mathcal H}^2 + 6 x_{\mathcal H}x_\sigma + x_\sigma^2\bigr)}
{\bigl(s-M_{Z^\prime}^2\bigr)^2}\,,
\\[2mm]
\sigma_{ \nu_\ell \bar\nu_\ell \to N_1 N_1}
&\simeq
\frac{g_L^4\,x_\sigma^2}{384\pi}\,
\frac{s\,(2x_{\mathcal H}+x_\sigma)^2}
{\bigl(s-M_{Z^\prime}^2\bigr)^2}\,,
\\[2mm]
\sigma_{h_1 h_1 \to N_1 N_1 }
&\simeq
\frac{M_{h_1}^4 M_{N_1}^2}{4\pi\,v_\sigma^4}\,
\frac{\sqrt{s} }{\bigl(s-M_{h_2}^2\bigr)^2\sqrt{s-4 M_{h_1}^2}}\,
\,,\label{eq:sigma_h1h1}
\\[2mm]
\sigma_{q_u \bar q_u \to N_1 N_1}
&\simeq
\frac{g_L^4\,x_\sigma^2}{10368 \pi}\,
\frac{s(34 x_{\mathcal{H}}^2 + 19 x_{\mathcal{H}} x_\sigma + x_\sigma^2)}{\bigl(s-M_{Z^\prime}^2\bigr)^2}\,,
\\[2mm]
\sigma_{q_d \bar q_d \to N_1 N_1 }
&\simeq
\frac{g_L^4\,x_\sigma^2}{10368\pi}\,
\frac{s(10 x_{\mathcal{H}}^2 - 2 x_{\mathcal{H}} x_\sigma + x_\sigma^2)}{\bigl(s-M_{Z^\prime}^2\bigr)^2}\,\,.
\end{align}
Here, $\ell$ denotes a charged lepton, $\nu_\ell$ the corresponding active neutrino flavor, 
and $q_{u,d}$ up- and down-type quarks with masses $m_{q_{u,d}}$. The phase-space factors in final states with massive particles are adequate approximations. Additional velocity-suppressed terms are negligible 
in the limit considered. 

Note that production initiated by SM gauge bosons ($ZZ$, $W^+W^-$) 
is strongly suppressed in the freeze-in regime because the relevant amplitudes 
are proportional to the scalar mixing angle. In particular, for $h_2$-mediated 
production the coupling to SM gauge bosons inherits a mixing suppression $\propto \sin\alpha$, 
implying the cross sections scaling as $\propto \sin^2\alpha$ (and in several 
channels as $\propto \sin^2 2\alpha$ depending on the interference structure). 
Since the phenomenologically viable region for freeze-in satisfies $|\alpha|\ll 1$, 
these gauge boson contributions rare several orders of magnitude smaller than the dominant channels listed and can be neglected.

\section{Phase-space distribution at the resonance}
\label{app:fq-resonance}
We show that the \textit{universal} freeze-in phase-space distribution~\cite{DEramo:2020gpr},
\begin{equation}
    f(q) \propto \frac{e^{-q}}{\sqrt{q}}\,,
    \label{eq:fq-universal}
\end{equation}
for DM produced via $2\to 2$ scattering is a good approximation near resonance.

The phase-space distribution can be written as \cite{Feiteira:2026qme}
\begin{equation}
    f(q) =
    \frac{1}{q}
    \int^{T_{\rm RH}}_{0}
    \frac{dT}{T^2}\,
    \frac{C(q,T)}{H(T)} \,,
 \label{eq:fq}   
\end{equation}
where $q=p/T$ is the comoving momentum. The collision term has the structure~\cite{DEramo:2020gpr},
\begin{equation}
    C(q,T) \propto
    \frac{e^{-q}}{q}
    \int_{s_{\rm min}}^{\infty} ds\, |\mathcal{M}|_{h_1 h_1 \leftrightarrow N_1 N_1}
    \frac{\sqrt{s-4M_{h_1}^2}}
    {s\sqrt{s-4M_{N_1}^2}}
    \exp\left[-\frac{s}{4qT^2}\right] ,
\end{equation}
with $s_{\rm min}=4\max(M_{h_1}^2,M_{N_1}^2)$. 
On resonance, 
\begin{equation}\label{eq:Amp_threshold}
    |\mathcal{M}|_{h_1 h_1 \leftrightarrow N_1 N_1} \propto \frac{s}{M_{h_2}^2 \Gamma_{h_2}^2}\,,
\end{equation}
and
\begin{equation}\label{eq:Threshold_cond}
    s \simeq M_{h_2}^2\,,
    \qquad
    M_{h_2} \simeq 2M_{h_1}\,.
\end{equation}
The integral over $s$ is dominated by a narrow interval of width
\begin{equation}
    \Delta s \sim M_{h_2}\Gamma_{h_2}\,,
\end{equation}
and the dependence on $q$ is mainly controlled  by the exponential factor,
\begin{equation}
    \exp\left[-q-\frac{s}{4qT^2}\right] .
\end{equation}
Thus, near the resonance, and assuming $M_{N_1} \ll M_{h_1}$, the distribution can be approximated as 
\begin{equation}
    f(q) \propto
    \frac{e^{-q}}{q^2}
    \int_0^{T_{\rm RH}} dT\,
    \frac{1}{T^2 H(T)}
    \int_{s_{\rm min}}^{M_{h_2}^2+\Delta s} ds\,
    \frac{1}{M_{h_2}^2\Gamma_{h_2}^2}
    \sqrt{1-\frac{4M_{h_1}^2}{s}}\,
    \exp\left[-\frac{s}{4qT^2}\right]\,.
\end{equation}
During radiation domination, $H(T)\propto T^2$, and the temperature
integral is
\begin{equation}
    I_T(q,s)
    \equiv
    \int_0^{T_{\rm RH}}
    \frac{dT}{T^4}\,
    \exp\left[-\frac{s}{4qT^2}\right] .
\end{equation}
For a sufficiently large reheating temperature 
with
\begin{equation}
    \frac{s}{4qT_{\rm RH}^2}\ll 1 \,,
\end{equation}
we obtain
\begin{equation}\label{eq:I_T_result}
    I_T(q,s)
    =
    \frac{\sqrt{\pi}}{4}
    \left(\frac{4q}{s}\right)^{3/2}
    +
    \mathcal{O}\!\left(\frac{s}{qT_{\rm RH}^2}\right).
\end{equation}
Substituting this into Eq.~(\ref{eq:fq}), we find
\begin{equation}
    f(q)
    \propto
    \frac{e^{-q}}{q^2}\,
    q^{3/2}
    \int_{s_{\rm min}}^{M_{h_2}^2+\Delta s} ds\,
    \frac{1}{s^{3/2}}\,
    \frac{1}{M_{h_2}^2\Gamma_{h_2}^2}
    \sqrt{1-\frac{4M_{h_1}^2}{s}}\,,
\end{equation}
so that the momentum dependence is simply
\begin{equation} \label{eq:PSD_resonance}
    f(q) \propto \frac{e^{-q}}{\sqrt{q}}\,.
\end{equation}

This shows analytically that the resonant enhancement affects the normalization of the distribution, but not its $q$ dependence. The reason is that the resonance makes $C(q,T)$ depend on a narrow range of $s$, while the $q$-dependence on the Boltzmann factor and radiation-dominated temperature integral remains unchanged.  For the benchmark of Eq.~\eqref{eq:benchmark-7keV}, we have numerically verified that the resulting distribution is well approximated by $f(q)\propto e^{-q}/\sqrt{q}$. 

It is important to note that Eq.~\eqref{eq:PSD_resonance} does not hold universally on resonance for DM produced via freeze-in or freeze-out. If $M_{h_2}\sim T_{\rm RH}$, the $q$-dependence of the phase-space distribution may be modified, as $I_T(q,s)$ 
cannot be evaluated in the limit $s/(4qT_{\rm RH}^2)\ll 1$. However, this is not the regime considered here as $T_{\rm RH}\gg M_{h_2}$.

\newpage
\bibliographystyle{JHEP}
\bibliography{Refs.bib}

\providecommand{\href}[2]{#2}\begingroup\raggedright\begin{thebibliography}{10}

\bibitem{Boyarsky:2018tvu}
A.~Boyarsky, M.~Drewes, T.~Lasserre, S.~Mertens, and O.~Ruchayskiy, {\it {Sterile neutrino Dark Matter}},  {\em Prog. Part. Nucl. Phys.} {\bf 104} (2019) 1--45.

\bibitem{Pal:1981rm}
P.~B. Pal and L.~Wolfenstein, {\it {Radiative Decays of Massive Neutrinos}},  {\em Phys. Rev. D} {\bf 25} (1982) 766.

\bibitem{Barret:2018qft}
D.~Barret {\em et~al.}, {\it {The Athena X-ray Integral Field Unit}},  {\em Proc. SPIE Int. Soc. Opt. Eng.} {\bf 10699} (2018) 106991G, [\href{https://arxiv.org/abs/1807.06092}{{\tt 1807.06092}}].

\bibitem{eROSITA:2012lfj}
{\bf eROSITA} Collaboration, A.~Merloni {\em et~al.}, {\it {eROSITA Science Book: Mapping the Structure of the Energetic Universe}},  \href{https://arxiv.org/abs/1209.3114}{{\tt 1209.3114}}.

\bibitem{XRISMScienceTeam:2020rvx}
{\bf XRISM Science Team} Collaboration, {\it {Science with the X-ray Imaging and Spectroscopy Mission (XRISM)}},  \href{https://arxiv.org/abs/2003.04962}{{\tt 2003.04962}}.

\bibitem{Bulbul:2014sua}
E.~Bulbul, M.~Markevitch, A.~Foster, R.~K. Smith, M.~Loewenstein, and S.~W. Randall, {\it {Detection of An Unidentified Emission Line in the Stacked X-ray spectrum of Galaxy Clusters}},  {\em Astrophys. J.} {\bf 789} (2014) 13, [\href{https://arxiv.org/abs/1402.2301}{{\tt 1402.2301}}].

\bibitem{Boyarsky:2014jta}
A.~Boyarsky, O.~Ruchayskiy, D.~Iakubovskyi, and J.~Franse, {\it {Unidentified Line in X-Ray Spectra of the Andromeda Galaxy and Perseus Galaxy Cluster}},  {\em Phys. Rev. Lett.} {\bf 113} (2014) 251301, [\href{https://arxiv.org/abs/1402.4119}{{\tt 1402.4119}}].

\bibitem{Stiele_2011}
H.~Stiele, W.~Pietsch, F.~Haberl, D.~Hatzidimitriou, R.~Barnard, B.~F. Williams, A.~K.~H. Kong, and U.~Kolb, {\it The deepxmm-newtonsurvey of m31},  {\em Astronomy \& Astrophysics} {\bf 534} (Oct., 2011) A55.

\bibitem{Dodelson:1993je}
S.~Dodelson and L.~M. Widrow, {\it {Sterile-neutrinos as dark matter}},  {\em Phys. Rev. Lett.} {\bf 72} (1994) 17--20, [\href{https://arxiv.org/abs/hep-ph/9303287}{{\tt hep-ph/9303287}}].

\bibitem{Dolgov:2000ew}
A.~D. Dolgov and S.~H. Hansen, {\it {Massive sterile neutrinos as warm dark matter}},  {\em Astropart. Phys.} {\bf 16} (2002) 339--344, [\href{https://arxiv.org/abs/hep-ph/0009083}{{\tt hep-ph/0009083}}].

\bibitem{Yuksel:2007xh}
H.~Yuksel, J.~F. Beacom, and C.~R. Watson, {\it {Strong Upper Limits on Sterile Neutrino Warm Dark Matter}},  {\em Phys. Rev. Lett.} {\bf 101} (2008) 121301, [\href{https://arxiv.org/abs/0706.4084}{{\tt 0706.4084}}].

\bibitem{Boyarsky:2007ge}
A.~Boyarsky, D.~Malyshev, A.~Neronov, and O.~Ruchayskiy, {\it {Constraining DM properties with SPI}},  {\em Mon. Not. Roy. Astron. Soc.} {\bf 387} (2008) 1345, [\href{https://arxiv.org/abs/0710.4922}{{\tt 0710.4922}}].

\bibitem{Loewenstein:2009cm}
M.~Loewenstein and A.~Kusenko, {\it {Dark Matter Search Using Chandra Observations of Willman 1, and a Spectral Feature Consistent with a Decay Line of a 5 keV Sterile Neutrino}},  {\em Astrophys. J.} {\bf 714} (2010) 652--662, [\href{https://arxiv.org/abs/0912.0552}{{\tt 0912.0552}}].

\bibitem{Horiuchi:2013noa}
S.~Horiuchi, P.~J. Humphrey, J.~Onorbe, K.~N. Abazajian, M.~Kaplinghat, and S.~Garrison-Kimmel, {\it {Sterile neutrino dark matter bounds from galaxies of the Local Group}},  {\em Phys. Rev. D} {\bf 89} (2014), no.~2 025017, [\href{https://arxiv.org/abs/1311.0282}{{\tt 1311.0282}}].

\bibitem{Neronov:2016wdd}
A.~Neronov, D.~Malyshev, and D.~Eckert, {\it {Decaying dark matter search with NuSTAR deep sky observations}},  {\em Phys. Rev. D} {\bf 94} (2016), no.~12 123504, [\href{https://arxiv.org/abs/1607.07328}{{\tt 1607.07328}}].

\bibitem{Perez:2016tcq}
K.~Perez, K.~C.~Y. Ng, J.~F. Beacom, C.~Hersh, S.~Horiuchi, and R.~Krivonos, {\it {Almost closing the {\ensuremath{\nu}}MSM sterile neutrino dark matter window with NuSTAR}},  {\em Phys. Rev. D} {\bf 95} (2017), no.~12 123002, [\href{https://arxiv.org/abs/1609.00667}{{\tt 1609.00667}}].

\bibitem{Roach:2019ctw}
B.~M. Roach, K.~C.~Y. Ng, K.~Perez, J.~F. Beacom, S.~Horiuchi, R.~Krivonos, and D.~R. Wik, {\it {NuSTAR Tests of Sterile-Neutrino Dark Matter: New Galactic Bulge Observations and Combined Impact}},  {\em Phys. Rev. D} {\bf 101} (2020), no.~10 103011, [\href{https://arxiv.org/abs/1908.09037}{{\tt 1908.09037}}].

\bibitem{Foster:2021ngm}
J.~W. Foster, M.~Kongsore, C.~Dessert, Y.~Park, N.~L. Rodd, K.~Cranmer, and B.~R. Safdi, {\it {Deep Search for Decaying Dark Matter with XMM-Newton Blank-Sky Observations}},  {\em Phys. Rev. Lett.} {\bf 127} (2021), no.~5 051101, [\href{https://arxiv.org/abs/2102.02207}{{\tt 2102.02207}}].

\bibitem{Irsic:2017ixq}
V.~Ir{\v{s}}i{\v{c}} {\em et~al.}, {\it {New Constraints on the free-streaming of warm dark matter from intermediate and small scale Lyman-$\alpha$ forest data}},  {\em Phys. Rev. D} {\bf 96} (2017), no.~2 023522, [\href{https://arxiv.org/abs/1702.01764}{{\tt 1702.01764}}].

\bibitem{Abazajian:2017tcc}
K.~N. Abazajian, {\it {Sterile neutrinos in cosmology}},  {\em Phys. Rept.} {\bf 711-712} (2017) 1--28, [\href{https://arxiv.org/abs/1705.01837}{{\tt 1705.01837}}].

\bibitem{Baur:2017stq}
J.~Baur, N.~Palanque-Delabrouille, C.~Yeche, A.~Boyarsky, O.~Ruchayskiy, E.~Armengaud, and J.~Lesgourgues, {\it {Constraints from Ly-$\alpha$ forests on non-thermal dark matter including resonantly-produced sterile neutrinos}},  {\em JCAP} {\bf 12} (2017) 013, [\href{https://arxiv.org/abs/1706.03118}{{\tt 1706.03118}}].

\bibitem{Boyarsky:2008ju}
A.~Boyarsky, O.~Ruchayskiy, and D.~Iakubovskyi, {\it {A Lower bound on the mass of Dark Matter particles}},  {\em JCAP} {\bf 03} (2009) 005, [\href{https://arxiv.org/abs/0808.3902}{{\tt 0808.3902}}].

\bibitem{Boyarsky:2008xj}
A.~Boyarsky, J.~Lesgourgues, O.~Ruchayskiy, and M.~Viel, {\it {Lyman-alpha constraints on warm and on warm-plus-cold dark matter models}},  {\em JCAP} {\bf 05} (2009) 012, [\href{https://arxiv.org/abs/0812.0010}{{\tt 0812.0010}}].

\bibitem{Gilman:2025fhy}
D.~Gilman {\em et~al.}, {\it {JWST lensed quasar dark matter survey IV: Stringent warm dark matter constraints from the joint reconstruction of extended lensed arcs and quasar flux ratios}},  \href{https://arxiv.org/abs/2511.07513}{{\tt 2511.07513}}.

\bibitem{Keeley:2025oig}
R.~E. Keeley {\em et~al.}, {\it {JWST Lensed Quasar Dark Matter Survey III: Dark Matter Sensitive Flux Ratios and Warm Dark Matter Constraint from the Full Sample}},  \href{https://arxiv.org/abs/2511.07765}{{\tt 2511.07765}}.

\bibitem{Keeley:2024brx}
R.~E. Keeley {\em et~al.}, {\it {JWST lensed quasar dark matter survey {\textendash} II. Strongest gravitational lensing limit on the dark matter free streaming length to date}},  {\em Mon. Not. Roy. Astron. Soc.} {\bf 535} (2024), no.~2 1652--1671, [\href{https://arxiv.org/abs/2405.01620}{{\tt 2405.01620}}].

\bibitem{Gilman:2026uvq}
D.~Gilman {\em et~al.}, {\it {The free-streaming length of dark matter from JWST observations of 28 strong gravitational lenses}},  \href{https://arxiv.org/abs/2606.05277}{{\tt 2606.05277}}.

\bibitem{Ruchayskiy:2015onc}
O.~Ruchayskiy, A.~Boyarsky, D.~Iakubovskyi, E.~Bulbul, D.~Eckert, J.~Franse, D.~Malyshev, M.~Markevitch, and A.~Neronov, {\it {Searching for decaying dark matter in deep XMM{\textendash}Newton observation of the Draco dwarf spheroidal}},  {\em Mon. Not. Roy. Astron. Soc.} {\bf 460} (2016), no.~2 1390--1398, [\href{https://arxiv.org/abs/1512.07217}{{\tt 1512.07217}}].

\bibitem{Urban:2014yda}
O.~Urban, N.~Werner, S.~W. Allen, A.~Simionescu, J.~S. Kaastra, and L.~E. Strigari, {\it {A Suzaku Search for Dark Matter Emission Lines in the X-ray Brightest Galaxy Clusters}},  {\em Mon. Not. Roy. Astron. Soc.} {\bf 451} (2015), no.~3 2447--2461, [\href{https://arxiv.org/abs/1411.0050}{{\tt 1411.0050}}].

\bibitem{Franse:2016dln}
J.~Franse {\em et~al.}, {\it {Radial Profile of the 3.55 keV line out to $R_{200}$ in the Perseus Cluster}},  {\em Astrophys. J.} {\bf 829} (2016), no.~2 124, [\href{https://arxiv.org/abs/1604.01759}{{\tt 1604.01759}}].

\bibitem{Cappelluti:2017ywp}
N.~Cappelluti, E.~Bulbul, A.~Foster, P.~Natarajan, M.~C. Urry, M.~W. Bautz, F.~Civano, E.~Miller, and R.~K. Smith, {\it {Searching for the 3.5 keV Line in the Deep Fields with Chandra: the 10 Ms observations}},  {\em Astrophys. J.} {\bf 854} (2018), no.~2 179, [\href{https://arxiv.org/abs/1701.07932}{{\tt 1701.07932}}].

\bibitem{XRISM:2025lzv}
{\bf XRISM} Collaboration, M.~Audard {\em et~al.}, {\it {XRISM Constraints on Unidentified X-Ray Emission Lines, Including the 3.5 keV Line, in the Stacked Spectrum of 10 Galaxy Clusters}},  {\em Astrophys. J. Lett.} {\bf 994} (2025), no.~1 L28, [\href{https://arxiv.org/abs/2510.24560}{{\tt 2510.24560}}].

\bibitem{Dessert:2023fen}
C.~Dessert, J.~W. Foster, Y.~Park, and B.~R. Safdi, {\it {Was There a 3.5 keV Line?}},  {\em Astrophys. J.} {\bf 964} (2024), no.~2 185, [\href{https://arxiv.org/abs/2309.03254}{{\tt 2309.03254}}].

\bibitem{Oda:2015gna}
S.~Oda, N.~Okada, and D.-s. Takahashi, {\it {Classically conformal U(1)' extended standard model and Higgs vacuum stability}},  {\em Phys. Rev. D} {\bf 92} (2015), no.~1 015026, [\href{https://arxiv.org/abs/1504.06291}{{\tt 1504.06291}}].

\bibitem{Porredon:2025xxv}
A.~Porredon {\em et~al.}, {\it {DESI-DR1 $3 \times 2$-pt analysis: consistent cosmology across weak lensing surveys}},  \href{https://arxiv.org/abs/2512.15960}{{\tt 2512.15960}}.

\bibitem{Semenaite:2025ohg}
A.~Semenaite {\em et~al.}, {\it {Joint cosmological fits to DESI-DR1 full-shape clustering and weak gravitational lensing in configuration space}},  \href{https://arxiv.org/abs/2512.15961}{{\tt 2512.15961}}.

\bibitem{Lange:2025rsx}
J.~U. Lange {\em et~al.}, {\it {Cosmological Constraints from Full-Scale Clustering and Galaxy-Galaxy Lensing with DESI DR1}},  \href{https://arxiv.org/abs/2512.15962}{{\tt 2512.15962}}.

\bibitem{Abdalla:2022yfr}
E.~Abdalla {\em et~al.}, {\it {Cosmology intertwined: A review of the particle physics, astrophysics, and cosmology associated with the cosmological tensions and anomalies}},  {\em JHEAp} {\bf 34} (2022) 49--211, [\href{https://arxiv.org/abs/2203.06142}{{\tt 2203.06142}}].

\bibitem{Fuss:2024dam}
L.~Fu{\ss}, M.~Garny, and A.~Ibarra, {\it {Minimal decaying dark matter: from cosmological tensions to neutrino signatures}},  {\em JCAP} {\bf 01} (2025) 055, [\href{https://arxiv.org/abs/2403.15543}{{\tt 2403.15543}}].

\bibitem{Astros:2025aan}
M.~D. Astros, L.~Gr{\'a}f, and S.~Vogl, {\it {Less structure on $8$ Mpc scales from decaying sterile neutrino dark matter}},  \href{https://arxiv.org/abs/2511.22638}{{\tt 2511.22638}}.

\bibitem{KM3NeT:2025npi}
{\bf KM3NeT} Collaboration, S.~Aiello {\em et~al.}, {\it {Observation of an ultra-high-energy cosmic neutrino with KM3NeT}},  {\em Nature} {\bf 638} (2025), no.~8050 376--382. [Erratum: Nature 640, E3 (2025)].

\bibitem{Kohri:2025bsn}
K.~Kohri, P.~K. Paul, and N.~Sahu, {\it {Super heavy dark matter origin of the PeV neutrino event: KM3-230213A}},  \href{https://arxiv.org/abs/2503.04464}{{\tt 2503.04464}}.

\bibitem{Chikashige:1980qk}
Y.~Chikashige, R.~N. Mohapatra, and R.~D. Peccei, {\it {Spontaneously Broken Lepton Number and Cosmological Constraints on the Neutrino Mass Spectrum}},  {\em Phys. Rev. Lett.} {\bf 45} (1980) 1926.

\bibitem{Chikashige:1980ui}
Y.~Chikashige, R.~N. Mohapatra, and R.~D. Peccei, {\it {Are There Real Goldstone Bosons Associated with Broken Lepton Number?}},  {\em Phys. Lett. B} {\bf 98} (1981) 265--268.

\bibitem{Gelmini:1980re}
G.~B. Gelmini and M.~Roncadelli, {\it {Left-Handed Neutrino Mass Scale and Spontaneously Broken Lepton Number}},  {\em Phys. Lett. B} {\bf 99} (1981) 411--415.

\bibitem{Goncalves:2024lrk}
J.~Gon{\c{c}}alves, D.~Marfatia, A.~P. Morais, and R.~Pasechnik, {\it {Gravitational waves from supercooled phase transitions in conformal Majoron models of neutrino mass}},  {\em JHEP} {\bf 02} (2025) 110, [\href{https://arxiv.org/abs/2412.02645}{{\tt 2412.02645}}].

\bibitem{Casas:2001sr}
J.~A. Casas and A.~Ibarra, {\it {Oscillating neutrinos and $\mu \to e, \gamma$}},  {\em Nucl. Phys. B} {\bf 618} (2001) 171--204.

\bibitem{Cordero-Carrion:2019qtu}
I.~Cordero-Carri\'on, M.~Hirsch, and A.~Vicente, {\it {General parametrization of Majorana neutrino mass models}},  {\em Phys. Rev. D} {\bf 101} (2020), no.~7 075032.

\bibitem{Esteban:2024eli}
I.~Esteban, M.~C. Gonzalez-Garcia, M.~Maltoni, I.~Martinez-Soler, J.~P. Pinheiro, and T.~Schwetz, {\it {NuFit-6.0: updated global analysis of three-flavor neutrino oscillations}},  {\em JHEP} {\bf 12} (2024) 216, [\href{https://arxiv.org/abs/2410.05380}{{\tt 2410.05380}}].

\bibitem{Planck:2018vyg}
{\bf Planck} Collaboration, N.~Aghanim {\em et~al.}, {\it {Planck 2018 results. VI. Cosmological parameters}},  {\em Astron. Astrophys.} {\bf 641} (2020) A6, [\href{https://arxiv.org/abs/1807.06209}{{\tt 1807.06209}}]. [Erratum: Astron.Astrophys. 652, C4 (2021)].

\bibitem{ATLAS:2022vkf}
{\bf ATLAS} Collaboration, G.~Aad {\em et~al.}, {\it {A detailed map of Higgs boson interactions by the ATLAS experiment ten years after the discovery}},  {\em Nature} {\bf 607} (2022), no.~7917 52--59, [\href{https://arxiv.org/abs/2207.00092}{{\tt 2207.00092}}]. [Erratum: Nature 612, E24 (2022)].

\bibitem{ATLAS:2020lks}
{\bf ATLAS} Collaboration, G.~Aad {\em et~al.}, {\it {Search for $ t\overline{t} $ resonances in fully hadronic final states in $pp$ collisions at $ \sqrt{s} $ = 13 TeV with the ATLAS detector}},  {\em JHEP} {\bf 10} (2020) 061, [\href{https://arxiv.org/abs/2005.05138}{{\tt 2005.05138}}].

\bibitem{ATLAS:2020tre}
{\bf ATLAS} Collaboration, G.~Aad {\em et~al.}, {\it {Search for lepton-flavour violation in high-mass dilepton final states using 139 fb$^{-1}$ of pp collisions at $ \sqrt{s} $ = 13 TeV with the ATLAS detector}},  {\em JHEP} {\bf 23} (2020) 082, [\href{https://arxiv.org/abs/2307.08567}{{\tt 2307.08567}}].

\bibitem{CMS:2024vhy}
{\bf CMS} Collaboration, A.~Hayrapetyan {\em et~al.}, {\it {Search for a Neutral Gauge Boson with Nonuniversal Fermion Couplings in Vector Boson Fusion Processes in Proton-Proton Collisions at s=13{\,}{\,}TeV}},  {\em Phys. Rev. Lett.} {\bf 135} (2025), no.~6 061803, [\href{https://arxiv.org/abs/2412.19261}{{\tt 2412.19261}}].

\bibitem{Khoze:2014xha}
V.~V. Khoze, C.~McCabe, and G.~Ro, {\it {Higgs vacuum stability from the dark matter portal}},  {\em JHEP} {\bf 08} (2014) 026, [\href{https://arxiv.org/abs/1403.4953}{{\tt 1403.4953}}].

\bibitem{Helo:2010cw}
J.~C. Helo, S.~Kovalenko, and I.~Schmidt, {\it {Sterile neutrinos in lepton number and lepton flavor violating decays}},  {\em Nucl. Phys. B} {\bf 853} (2011) 80--104, [\href{https://arxiv.org/abs/1005.1607}{{\tt 1005.1607}}].

\bibitem{Barger:1995ty}
V.~D. Barger, R.~J.~N. Phillips, and S.~Sarkar, {\it {Remarks on the KARMEN anomaly}},  {\em Phys. Lett. B} {\bf 352} (1995) 365--371, [\href{https://arxiv.org/abs/hep-ph/9503295}{{\tt hep-ph/9503295}}]. [Erratum: Phys.Lett.B 356, 617--617 (1995)].

\bibitem{Drewes:2016upu}
M.~Drewes {\em et~al.}, {\it {A White Paper on keV Sterile Neutrino Dark Matter}},  {\em JCAP} {\bf 01} (2017) 025, [\href{https://arxiv.org/abs/1602.04816}{{\tt 1602.04816}}].

\bibitem{Alwall:2014hca}
J.~Alwall, R.~Frederix, S.~Frixione, V.~Hirschi, F.~Maltoni, O.~Mattelaer, H.~S. Shao, T.~Stelzer, P.~Torrielli, and M.~Zaro, {\it {The automated computation of tree-level and next-to-leading order differential cross sections, and their matching to parton shower simulations}},  {\em JHEP} {\bf 07} (2014) 079, [\href{https://arxiv.org/abs/1405.0301}{{\tt 1405.0301}}].

\bibitem{Hall:2009bx}
L.~J. Hall, K.~Jedamzik, J.~March-Russell, and S.~M. West, {\it {Freeze-In Production of FIMP Dark Matter}},  {\em JHEP} {\bf 03} (2010) 080, [\href{https://arxiv.org/abs/0911.1120}{{\tt 0911.1120}}].

\bibitem{Gondolo:1990dk}
P.~Gondolo and G.~Gelmini, {\it {Cosmic abundances of stable particles: Improved analysis}},  {\em Nucl. Phys. B} {\bf 360} (1991) 145--179.

\bibitem{ParticleDataGroup:2024cfk}
{\bf Particle Data Group} Collaboration, S.~Navas {\em et~al.}, {\it {Review of particle physics}},  {\em Phys. Rev. D} {\bf 110} (2024), no.~3 030001.

\bibitem{Staub:2013tta}
F.~Staub, {\it {SARAH 4 : A tool for (not only SUSY) model builders}},  {\em Comput. Phys. Commun.} {\bf 185} (2014) 1773--1790, [\href{https://arxiv.org/abs/1309.7223}{{\tt 1309.7223}}].

\bibitem{Belyaev:2012qa}
A.~Belyaev, N.~D. Christensen, and A.~Pukhov, {\it {CalcHEP 3.4 for collider physics within and beyond the Standard Model}},  {\em Comput. Phys. Commun.} {\bf 184} (2013) 1729--1769, [\href{https://arxiv.org/abs/1207.6082}{{\tt 1207.6082}}].

\bibitem{Abazajian:2005gj}
K.~Abazajian, {\it {Production and evolution of perturbations of sterile neutrino dark matter}},  {\em Phys. Rev. D} {\bf 73} (2006) 063506, [\href{https://arxiv.org/abs/astro-ph/0511630}{{\tt astro-ph/0511630}}].

\bibitem{Patwardhan:2015kga}
A.~V. Patwardhan, G.~M. Fuller, C.~T. Kishimoto, and A.~Kusenko, {\it {Diluted equilibrium sterile neutrino dark matter}},  {\em Phys. Rev. D} {\bf 92} (2015), no.~10 103509, [\href{https://arxiv.org/abs/1507.01977}{{\tt 1507.01977}}].

\bibitem{Garzilli:2019qki}
A.~Garzilli, A.~Magalich, O.~Ruchayskiy, and A.~Boyarsky, {\it {How to constrain warm dark matter with the Lyman-$\alpha$ forest}},  {\em Mon. Not. Roy. Astron. Soc.} {\bf 502} (2021), no.~2 2356--2363, [\href{https://arxiv.org/abs/1912.09397}{{\tt 1912.09397}}].

\bibitem{blas2011cosmic}
D.~Blas, J.~Lesgourgues, and T.~Tram, {\it The cosmic linear anisotropy solving system (class). part ii: Approximation schemes},  {\em Journal of Cosmology and Astroparticle Physics} {\bf 2011} (2011), no.~07 034.

\bibitem{DEramo:2020gpr}
F.~D'Eramo and A.~Lenoci, {\it {Lower mass bounds on FIMP dark matter produced via freeze-in}},  {\em JCAP} {\bf 10} (2021) 045, [\href{https://arxiv.org/abs/2012.01446}{{\tt 2012.01446}}].

\bibitem{Schneider:2016uqi}
A.~Schneider, {\it {Astrophysical constraints on resonantly produced sterile neutrino dark matter}},  {\em JCAP} {\bf 04} (2016) 059, [\href{https://arxiv.org/abs/1601.07553}{{\tt 1601.07553}}].

\bibitem{Murgia:2017lwo}
R.~Murgia, A.~Merle, M.~Viel, M.~Totzauer, and A.~Schneider, {\it {''Non-cold'' dark matter at small scales: a general approach}},  {\em JCAP} {\bf 11} (2017) 046, [\href{https://arxiv.org/abs/1704.07838}{{\tt 1704.07838}}].

\bibitem{Zelko:2022tgf}
I.~A. Zelko, T.~Treu, K.~N. Abazajian, D.~Gilman, A.~J. Benson, S.~Birrer, A.~M. Nierenberg, and A.~Kusenko, {\it {Constraints on Sterile Neutrino Models from Strong Gravitational Lensing, Milky~Way Satellites, and the Lyman-{\ensuremath{\alpha}} Forest}},  {\em Phys. Rev. Lett.} {\bf 129} (2022), no.~19 191301, [\href{https://arxiv.org/abs/2205.09777}{{\tt 2205.09777}}].

\bibitem{Feiteira:2026qme}
D.~Feiteira, O.~Lebedev, and V.~Oliveira, {\it {Warm dark matter from freeze-in at stronger coupling}},  {\em Phys. Lett. B} {\bf 877} (2026) 140486, [\href{https://arxiv.org/abs/2602.20242}{{\tt 2602.20242}}].

\end{thebibliography}\endgroup

\end{document}